# LASSO-ODE: A framework for mechanistic model identifiability and selection in disease transmission modeling

## Jiale Tan[1], Marisa Eisenberg[2]

jialetan@umich.edu, marisae@umich.edu
[1]Epidemiology and Scientific Computing, University of Michigan, Ann Arbor
[2]Epidemiology, Complex Systems, and Mathematics, University of Michigan, Ann Arbor

## Abstract

To be fully useful for public health practice, models for epidemic response must be able to do more than predict—it is also important to incorporate the mechanisms underlying transmission dynamics to enable policymakers and practitioners to be able to evaluate what-if scenarios and intervention options. However, most mechanistic models suffer from uncertainty in both the parameters (e.g., parameter unidentifiability) and the model structure itself, which can hinder both successful parameter estimation and model interpretation. To enable rapid development of interpretable and parsimonious mechanistic models, we use penalized regression and covariate selection methods to integrate parameter identifiability and model selection directly into the parameter estimation procedure for (in this case) traditional ordinary differential equation (ODE) models. For both simulated and real-world epidemiological data, we demonstrate that the LASSO-ODE framework is highly effective in selecting a parsimonious, identifiable model from larger, more realistic but potentially unidentifiable models, from realistically sparse data with only a single measured compartment and multiple latent (unobserved) variables. While we focus on epidemic models in this paper as a case study, these same approaches are applicable to a wide range of application areas that are faced with relatively sparse data but a need for realistic mechanistic models (e.g. mathematical oncology and mathematical biology more broadly). Additionally, the cross-validation techniques designed for time series data introduced in our study can be used across a range of time series analysis and modeling approaches.

## 1 Introduction

Mechanistic models are essential for policymakers in public health for forecasting disease burden, comparing alternative intervention strategies, and informing decision-making.[1] However, mechanistic models often make significant assumptions about disease dynamics—for example, the classical Susceptible-Infectious-Recovered (SIR) model makes numerous assumptions around the transmission pathway, random mixing, constant exponential rates of recovery, and other features.[2] Although these models are in principle clear and interpretable, once confronted with real data, they can often suffer from uncertainty in their parameters and model structure that can compromise the model's effectiveness in prediction and causal inference.[3,4]

Two key aspects of the uncertainty faced by mechanistic models include parameter unidentifiability and model misspecification (and the related question of model selection).[5] A model is identifiable if it has a unique set of parameters that can be estimated from a given data set.[6] If not, predictions and interpretations can be unreliable. Unidentifiability arises either from the inherent design of the model (structural unidentifiability) or from the limitations of the dataset (practical unidentifiability).[6] Structural identifiability is focused on the theoretical capacity to estimate model parameters from an ideal dataset without noise, while practical identifiability deals with real-world data quality and its effect on parameter estimation.[7] Addressing these issues is challenging but can be mitigated by enhancing data quality, imposing parameter constraints based on scientific knowledge, or restructuring the model to reduce the dimension of the parameter space. However, these measures require careful consideration based on the model and data characteristics, and there is a continued need for developing more sophisticated and generalized methods to deal with identifiability problems.[19,20] Existing methods for addressing identifiability issues include differential algebra methods[29], matrix-rank-based methods using sensitivity or observability matrices[31,32], profile likelihoods[30], and more. Each of these methods has distinct advantages and disadvantages in terms of applicability to a wide range of models, computational cost, and complexity of implementation. These approaches also typically require additional computation on top of the existing parameter estimation process and generally do not directly address model selection (though in some cases this can be inferred from the process of resolving the identifiability issue).

The related concept of model selection is the process of choosing the most appropriate model from a set of candidate models for a given dataset and research question.[8] This process involves comparing different models based on their performance, complexity, and interpretability, among other factors. A major goal of model selection is to find a balance between fitting the data well (or sometimes relatedly, having high predictive accuracy) while maintaining relative simplicity and avoiding overfitting (where the model is too complex and captures noise rather than the underlying pattern). A danger in the process of model selection for mechanistic models is that the selected model may not reflect the true underlying mechanism—rather just the simplest model that fits the data well.[18] Addressing this concern requires careful thought, domain knowledge, and also knowledge of how to interpret the fitted parameters (as the optimization algorithm does not use the biological or mechanistic meaning of the parameters, and in cases of model misspecification may use the model in unintended ways).

Given the challenges of identifiability and model selection in mechanistic models, machine learning offers a flexible alternative that can manage large datasets without needing a predefined model structure.[9,10] For example, tree-based models such as random forest and XG-boost, along with neural networks, do not assume a specific model structure, but rather learn it as part of the estimation process.[16,17] However, machine learning models often lack biological interpretability and may not forecast accurately in new settings. This poses a significant obstacle for public health decision-makers, who depend on a clear understanding of how various factors contribute to disease progression and the implications of potential interventions (what-if scenarios).[11,12]

Considering the limitations of both mechanistic and machine learning models, this study aims to establish a framework for selecting the most interpretable models or sub models from one or more mechanistic candidates, specifically focusing on ordinary differential equation (ODE) models. To achieve this, we integrate statistical penalized regression techniques by incorporating the LASSO penalty[13] into the cost function of various model candidates. This encourages sparsity and covariate selection (in this case the covariates meaning parameters or portions of models), helping us identify the most appropriate (sub)model. By applying the LASSO penalty to traditional ODE models in epidemic forecasting, our goal is to enable improved identifiability and model selection directly within the estimation process, allowing these issues to be addressed simultaneously.

LASSO and similar penalized regression approaches have been used for model/variable selection in statistical models for decades[28], where they are often used to find estimable subsets of covariates/variables for regression problems. However, these approaches have not caught on as thoroughly in the mechanistic modeling literature, even though issues of model selection and model identifiability/estimability are quite common in these models.[29] However, one area of the mechanistic modeling literature that does make use of LASSO and similar penalized regression methods is the area of data-driven model discovery using general function families. Brunton et al. developed a key framework in this area denoted SINDy (Sparse Identification of Nonlinear Dynamics) for discovering governing equations from large datasets.[25] The core idea of this approach is to construct the most likely governing equations without any prior knowledge, using a library of functional terms, the coefficients of which are estimated with sparse regression techniques. Egan et al. and others have since expanded this approach to develop a range of similar methods.[26]

However, this approach is fundamentally based on discovering the model "from scratch" rather than evaluating the identifiability or selection of pre-established mechanistic models. This is an important question that addresses the broader issue of model selection and model uncertainty—but it is fundamentally a more difficult and distinct question than the one often faced in epidemiology where the basic modeling framework (SIR) is usually known and one or more realistic candidate models are often already determined based on the disease biology. Additionally, previous work using SINDy and related methods for infectious disease modeling has shown mixed success, with some infectious disease patterns either not able to be fit or overfitted with unrealistic model structures.[27] This is likely in part due to some of the restrictions of most automated model-discovery methods around needing high-quality time series data, ideally for all variables in the model (e.g. susceptible and asymptomatically infected individuals, which are typically not measured), as well as the challenges of algorithmically selecting a model from complex time series data (when likely many different functional forms that may not be mechanistic may match the data well). These data restrictions are often not met in real-world epidemiological or biological data, which is frequently sparse, noisy, and measures only a handful of compartments (often only one, the number of infected individuals).

As noted above, in real-world scenarios in infectious disease epidemiology, we often have well-established model systems to describe the data, often based on the SIR framework. Thus the focus of this paper is somewhat different than that of SINDy: to select the appropriate model or submodel and identifiable parameters from these pre-existing candidates. Specifically, the **objective** of our study is to evaluate whether our approach can consistently select an identifiable model for a given data set from a set of candidate models and accurately estimate parameters by removing irrelevant ones. We assess our approach using a series of critical inquiries:

- **Q1)** Can our framework effectively mitigate overfitting?
- **Q2)** Is our framework capable of avoiding underfitting?
- **Q3)** Does our framework facilitate the identifiability of our model?
- **Q4)** Are we able to utilize our framework to choose an appropriate model from multiple model candidates?
- **Q5)** Is our model sufficiently interpretable to be useful to public health practitioners and policymakers?

# 2 LASSO-ODE framework

## 2.1 Overall framework

The basic idea for the methods proposed here is to use a LASSO penalized objective function (together with novel cross-validation methods) to select from amongst a set of realistic candidate models, where 1) we may have multiple models incorporated in parallel into a single objective function to select among, 2) some portions of the model(s) may be unidentifiable, which will mean that a submodel of the larger model may be selected (as some unidentifiable parameters will be reduced to zero), 3) some selected parameters may be chosen to be non-penalized (if there are portions of the model(s) that are known from the literature/data or important to maintain as part of the model). In doing so, we hope to obtain identifiable submodel(s) that fit the data well, enabling us to perform simultaneous parameter estimation, model selection, and reduction/reparameterization to an identifiable model.

The framework proposed in this study is comprised of four main modules: the Data module, the Objective function module, the Cross-validation module, and the Output. The data (either simulated or real-world data) and the objective function are both used by the Cross-Validation module, which runs both parameter estimation for the model parameters and determines the appropriate value of the sparsity penalty $\lambda$ via cross-validation. In this study, we use least squares for parameter estimation (but this portion of the objective function could be replaced with any other similar approach, e.g. the likelihood function). We have developed four different cross-validation methods to accommodate various data challenges, such as incomplete datasets, rapidly changing epidemic trends, or computational limitations. The availability of multiple methods allows us to adapt flexibly to different data-related scenarios, ensuring robustness in our estimation process. To better understand how our framework operates, Figure 1 illustrates an example of how LASSO-ODE might select the most relevant model and estimate the best parameters from two model candidates. Example code for the LASSO-ODE approach is available at https://github.com/epimath/lasso-ode, including multiple cross-validation methods and data sets (simulated and actual).

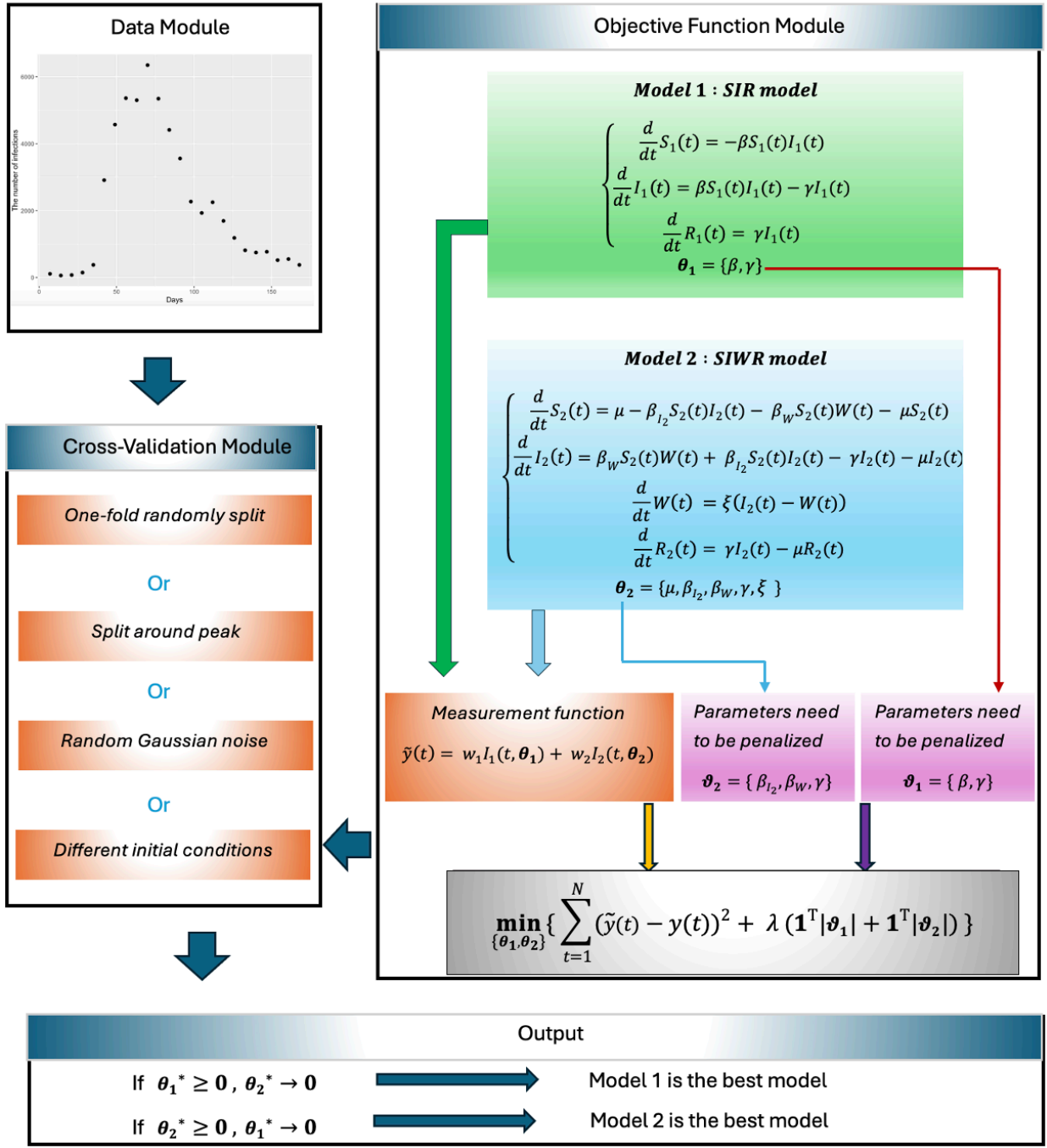

Figure 1. LASSO-ODE Framework Schematic Diagram and Example: We aim to identify the optimal model and parameters to fit the infection data, using two model candidates: SIR and SIWR. We note that the SIWR model candidate also includes two plausible submodels: the person-to-person only transmission model ($\beta_W = 0$, equivalent to the SIR model), and the waterborne-only transmission model ($\beta_{I_2} = 0$). LASSO-ODE uses the data and penalized regression cost function to estimate the model parameters and select the appropriate penalization. Based on which parameters are penalized to near zero, we can determine which model is selected based on the data and evaluate if the model has been simplified to facilitate identifiability.

### 2.1.1 Data module

We tested our method with two types of datasets—simulated (see details in Section 2.2) and then real-world (see details in Section 2.3)—to practically evaluate our framework's performance. Initially, we used simulated data to quantify the procedure's accuracy by directly comparing the estimated parameters to the actual parameters used to generate the data. Following this evaluation, we applied the validated framework to real-world data. Typically, measured data for infectious disease surveillance involves some measurement of the number of either new cases (incidence) or currently infectious individuals (prevalence). Here, we assumed the data used for LASSO-ODE consists of the number of infectious individuals at each time point.

The simulated data is produced using an ordinary differential equation (ODE) model, referred to as the "True model" and detailed in Table 1 in Section 2.2, along with a set of specific parameters, where the data is taken to be one simulation of the measurement function defined below. To ensure a comprehensive data set for training the model, the simulation's timeframe was extended until the full epidemic curve was captured. For the real-world data component, we used cholera time series data of weekly incident cases that were compiled in 2006 during a large epidemic centered in Luanda, Angola.[14]

### 2.1.2 Objective function module

Our objective function consists of the squared error between the predicted and actual number of infectious individuals, along with a LASSO penalty term applied to a subset of the parameters in the predictive model (depending on which parameters are candidates for elimination). The predicted number of infectious individuals is derived from the weighted sum of the infectious compartments in each predictive model. Specifically, if there is only one predictive model, the predicted infectious count is obtained solely from the infectious compartment(s) of that model. In a more general scenario with multiple potential model candidates, the predicted infectious count is formed by summing the infectious compartments across all model candidates with a weight assigned to each model. By including all potential model candidates in a single objective function, our framework can identify the most suitable model or sub-model and estimate its parameters. This method anticipates that parameters unrelated to the true or selected model will converge to zero—in particular, the weights in the objective function for non-selected models will be near-zero, indicating these models were not selected.

The LASSO penalty, commonly referred to as the L1 penalty, is a method employed in variable selection and regularization. Unlike the L2 penalty, which merely shrinks the coefficients without ever necessarily reaching zero, the LASSO penalty possesses the distinctive capability to reduce the coefficients of less significant variables to zero, thereby excluding them from the model. Therefore, it can help us simplify the model and then make our model more identifiable. The mathematical formulation of the LASSO penalty is incorporated into the objective function as follows—if we have K potential model candidates:

$$\frac{d}{dt}X_1(t) = f_1(X_1(t), \theta_1) \quad \textbf{\textit{Model 1}}$$

$$\frac{d}{dt}X_2(t) = f_2(X_2(t), \theta_2) \quad \textbf{\textit{Model 2}}$$

$$... \qquad\qquad ...$$

$$\frac{d}{dt}X_K(t) = f_K(X_K(t), \theta_K) \quad \textbf{\textit{Model K}}$$

Here, $\boldsymbol{X_i}(t)$ is a vector representing the $n_i$ compartment variables (disease states or other compartments $\boldsymbol{X_i}(t) = [x_{i1}(t), x_{i2}(t), \ldots, x_{in_i}(t)]$) at time $t$ for the $i$-th model, $\boldsymbol{\theta_i}$ is a vector that contains all the parameters for the $i$-th model, and $\boldsymbol{f_i}(\boldsymbol{X_i}(t), \boldsymbol{\theta_i})$ here is a vector representing $n_i$ rate equations the $i$-th model, expressing the interactions between the compartments and the parameters. For disease models, this can often be written as $\boldsymbol{A_i X_i} + g(\boldsymbol{X_i})$, where $\boldsymbol{A_i}$ is an $n_i \times n_i$ matrix representing linear transitions between compartments that do not involve interaction terms like products of the states. $g(\boldsymbol{X})$ is a vector representing nonlinear terms, which often include product terms like those seen in the transmission of diseases (e.g., $\beta x_{k1} x_{k2}$ for infection).

For each model, we select a measurement function $h_i$ that represents the observed quantity in the data—often this is a single compartment/variable, potentially multiplied by a parameter (e.g. $h_i = I$ would measure prevalence in a classic SIR model). We sum these different measurement functions for each model, with a weight $\omega_i$ to generate our overall measurement function:

$$\tilde{y}(t) = \sum_{i=1}^{K} \omega_i h_i(\boldsymbol{X_i}(t), \boldsymbol{\theta_i})$$

Each weight parameter $\omega_i$ indicates the extent to which $h_i(\boldsymbol{X_i}(t), \boldsymbol{\theta_i})$ contributes to the measurement function, with the constraint that the sum of the weights equals to 1 for ease of interpretation. Ideally, if the $m$-th model is the true model (or closest to it), then $\omega_{i=m} = 1$, and $\omega_{i \neq m} = 0$. If there is only one model candidate and we want to simplify it into a sub model, then this model's weight is fixed at 1, meaning we will use only the number of infections from this model as the measurement function.

Traditional ODE disease transmission models are structured around various disease statuses, each represented by a distinct compartment in the disease transmission model. For instance, the classical SIR model[24] includes three compartments: Susceptible, Infectious, and Recovered. Measurements for this model typically focus on the number of infectious individuals, as they are more likely to visit hospitals or clinics for treatment and thus are more likely to be monitored. Consequently, the data used in our framework is the number of infectious individuals at each time point. Therefore, the measurement function is defined as below:

$$\tilde{y}(t) = \sum_{i=1}^{K} \omega_i I_i(\boldsymbol{X_i}(t), \boldsymbol{\theta_i})$$

In this context, $I_i(\boldsymbol{X_i}(t), \boldsymbol{\theta_i})$ denotes the observed number of infections in the $i$-th model at time $t$, and $\omega_i$ is the weight parameter for the $i$-th model.

All together, our objective function is formed by:

$$\min_{\{\boldsymbol{\theta_1}, \boldsymbol{\theta_2} \ldots \boldsymbol{\theta_K}\}} \left\{ \sum_{t=1}^{N} \left( \tilde{y}(t) - y(t) \right)^2 + \lambda \sum_{i=1}^{K} \mathbf{1}^{\mathrm{T}} | \boldsymbol{\vartheta_i} | \right\}$$

Where $y(t)$ is the actual number of infections in the data (either simulated or real world), $\lambda$ is the regularization hyper parameter, $\boldsymbol{\vartheta_i}$ are the selected parameters that we need to penalize in $i$-th model, and $N$ is the number of time points in the data.

Parameter penalization in the objective function serves several purposes, including the selection of relevant features of the model. Penalization causes less important parameters to shrink towards zero, thereby implicitly removing these features from the model. Traditionally in LASSO regression, such penalization is applied uniformly across all parameters. However, when dealing with infectious disease models, certain biological parameters have previously been accurately estimated from various sources and exhibit minimal variation across populations, or may otherwise be well established to be a part of the mechanistic process. Consequently, these parameters are consistently relevant and should not be eliminated from our predictive models. Thus, we apply penalization selectively, targeting only those parameters that might justifiably be reduced to zero, to optimize the model effectively (although the method can also be applied to all parameters as well, as these known parameters will often be maintained as non-zero as they are important to the dynamics).

### 2.1.3 Cross-validation module

This module estimates the model parameters and weights using the Nelder-Mead method, and uses cross validation to test a series of values for $\lambda$ (the LASSO penalty scaling factor). To determine the values of $\lambda$ to test for cross-validation, we first ran a trial run of parameter estimation with $\lambda = 0$, and then used the magnitude of the sum of squared residuals to select a range of $\lambda$ values spanning one order of magnitude below to one order of magnitude above the size of the squared residuals term. Ultimately, we selected the one with the best predictive value (including potentially $\lambda = 0$ if no model reduction is needed). Cross-validation for time series data requires special consideration because standard cross-validation techniques like k-fold cross-validation can lead to incorrect conclusions due to the non-independence of temporal data. This is due to the temporal dependencies within time series data, meaning that observations are typically correlated with previous observations. Thus, data from future time points often should not be validated using previous time points (unless the series is time independent). Instead, blocking cross-validation methods[22] have been used for time series data. Blocking cross-validation is a method that ensures that for each split of the data set, the training set is always composed of data points that occurred prior to those in the validation set. We have refined the cross-validation process by introducing four distinct methods in addition to the conventional blocking cross-validation technique for our analyses, which we compare and test. These methods each ensure that the validation process respects the chronological order of the data and guards against the inadvertent use of future information to predict past observations. In the main text, we present only Algorithm 1 (One-fold randomly split) as it performed best in cross-validation across most scenarios and was considered the most intuitive. Algorithms 2 through 4 are included in the supplementary materials for scenarios where different ODE types might be more suitable if Algorithm 1 is inadequate.

---

**Algorithm 1: One-fold randomly split**

---

1: **Input:** Observed time series infections $\mathbf{I} = \{I_t\}_{t=1}^N$ , where $N$ here is the total number of time points.

       A range of $\boldsymbol{Lambda} = \boldsymbol{\lambda}$.

       Initial guess of parameters in the prediction model $\boldsymbol{\theta_0} = \{\theta_i\}_{i=1}^P$,

       where $P$ here is the total number of parameters in the model.

       The total number of simulations $C$.

2. **Output:** Corresponding optimal parameters $\overline{\boldsymbol{\theta}^*}$

3. **for** $\lambda$ in **Lambda, do**
4.     $k = 1$, where $k$ here is the counting number of the simulations
5.     **while** $(k < C)$, **do**
6.         Cutting_point $\sim$ U $(1, N)$
7.         $\mathbf{I_{train,k}} = \{I_t\}_{t=1}^{\text{Cutting\_point}}$
8.         $\mathbf{I_{test,k}} = \{I_t\}_{t=\text{Cutting\_point}+1}^{N}$
9.         $\boldsymbol{\theta_{k0}} = \boldsymbol{\theta_0} + \mathcal{N}(\boldsymbol{\theta_0},\ 0.2)$, here the variance 0.2 can be any positive value
10.        $\boldsymbol{\theta_{k,\lambda}} = \underset{\theta}{\operatorname{argmin}}\{\sum(\widetilde{\mathbf{I_{train,k}}} - \mathbf{I_{train,k}})^2 + \lambda \sum_j |\vartheta_j|\ \}$ given $\boldsymbol{\theta_{k0}}$, $I_{t=1}$, and $\lambda$
           where $\mathbf{I_{\widetilde{train,k}}}$ is the measurement function values using the training set,
              $\vartheta_j$ is the parameters which need to be penalized
11.        $\widetilde{\mathbf{I_{test,k}}} =$ The predicted number of infections given $\boldsymbol{\theta_{k,\lambda}}$, and $I_{t=\text{Cutting\_point}+1}$
12.        $\text{MSE}_{test,k} = \frac{1}{length(\mathbf{I_{test,k}})}\sum(\widetilde{\mathbf{I_{test,k}}} - \mathbf{I_{test,k}})^2$
13.        $k = k+1$
14.    **end while**
15.    $\boldsymbol{\theta_\lambda} = \{\boldsymbol{\theta_{k,\lambda}}\}_{k=1}^C$ is the collection of $\boldsymbol{\theta_{k,\lambda}}$ from each $k$th simulation under $\lambda$
16.    $\overline{\text{MSE}}_{\lambda,\boldsymbol{\theta_\lambda}} =$ mean (or median) of $\{\text{MSE}_{test,k}\}_{k=1}^C$ from $C$ times simulation under $\lambda$
17. **end for**
18. $\lambda^* = \underset{\lambda,\theta_\lambda}{\operatorname{argmin}}\{\ \overline{\text{MSE}}_{\lambda,\boldsymbol{\theta_\lambda}}\ \}$
19. $\boldsymbol{\theta^*}$ is the $\boldsymbol{\theta_\lambda}$ under $\lambda^*$
20. $\overline{\boldsymbol{\theta^*}} =$ median or mean of $\boldsymbol{\theta^*}$ from $C$ times simulation under $\lambda^*$
21. Return $\overline{\boldsymbol{\theta^*}}$ (you can also return $\boldsymbol{\theta^*}$ to see the entire distribution of $C$ times estimation)

---

In this approach, we simplify the blocking cross-validation approach and strategically partition our time series dataset into two contiguous segments with a single random split point. The initial segment, which extends from the initial time point (time 1) up to a cut-off point we randomly choose, serves as the training set for developing our model. The subsequent segment, which runs from that cut-off point to the end of the series, is designated for model validation. The rationale for utilizing a single validation fold lies in the premise that the most pertinent information—the key signals—predominantly reside within the first portion of the data (in the case of epidemic data, often during the initial rise, inflection, and peak in the epidemic).[15, 23] Introducing additional folds could potentially dilute these signals and introduce confusion, possibly leading to less reliable model performance. This method ensures that the model learns from historical data and predicts future trends, maintaining the chronological integrity of the dataset. However, there is some possibility that the random split may make either the test or training data uninformative.

## 2.2 Framework evaluation with simulated data

As described above, we next tested our framework by using a test/predictive model to estimate parameters using data simulated from a true model. This approach allows us to evaluate the accuracy of our framework by testing whether the framework is able to uncover the true parameters used to simulate the data. Additionally, even when using this method in actual applications, an initial simulation phase using simulated data from the model(s) under consideration may be useful to assess whether the model(s) are structurally identifiable.

We assessed our models using three distinct scenarios:
1. The true model is embedded in the predictive model (*embedded*)
2. The true model is identical to the predictive model (*identical*)
3. There are multiple predictive models whose outputs are weighted summed (to generate a single large model), and one of these alternative models is identical to the true model (*parallel*)

In each scenario, we generated simulated data using the specified parameters of the true model. Subsequently, we employed the predictive model to fit this simulated data, aiming to a) reduce the model to the correct predictive model as needed, and b) correctly infer the parameters of the predictive model.

Our examples for each scenario are as follows:

| | Prediction model | True model |
|---|---|---|
| Embedded Scenario | $SI_2R$ model:<br><br>$S' = -\beta_1 S I_1 - \beta_2 S I_2$<br>$I_1' = \beta_1 S I_1 - \gamma_1 I_1$<br>$I_2' = \beta_2 S I_2 - \gamma_2 I_2$<br>$R' = \gamma_1 I_1 + \gamma_2 I_2$<br><br>$\tilde{y} = I_1 + I_2$<br><br>Where $S$, $I_n$, and $R$ represent the fractions of the population that are susceptible, infectious of the $n$th type (n = 1, 2), and recovered, respectively. Parameters $\beta_n$ and $\gamma_n$ represent the rate of transmission and recovering from $n$th infection type, respectively. $\tilde{y}$ represents measured output. In addition, we made the initial conditions parameters to be estimated and penalized all parameters in the model. Note that the true model is embedded within the predictive model in this case (by selecting only one of the two $I$ compartments and setting the parameters for the other to zero). We penalized all parameters in this model.<br><br>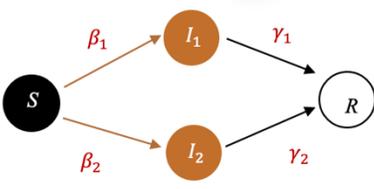 | $SIR$ model:<br><br>$S' = -\beta S I$<br>$I' = \beta S I - \gamma I$<br>$R' = \gamma I$<br><br>Where $S$, $I$ and $R$ represent the fractions of the population that are susceptible, infectious, and recovered, respectively. Parameters $\beta$ and $\gamma$ represent the rate of transmission and recovery.<br><br>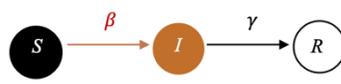 |

| | *SIWR* model[7,21]: | *SIR* model: |
|---|---|---|
| | $S' = \mu - \beta_I SI - \beta_W SW - \mu S$ | Variables and parameters are as |
| | $I' = \beta_W SW + \beta_I SI - \gamma I - \mu I$ | described above. |
| | $W' = \xi(I - W)$ | |
| | $R' = \gamma I - \mu R$ | |
| | | |
| | $\tilde{y} = I$ | |
| | | |
| | Where $S$, $I$, $W$ and $R$ represent the fractions of the population that are susceptible, infectious, proportional to the concentration of waterborne pathogen in the environment, and recovered, respectively. $\tilde{y}$ represents measured output. Parameter $\xi$ represents the decay rate of the pathogen in the water. Parameters $\beta_W$ and $\beta_I$ represent the rate of transmission from indirect water-person and direct person-person, respectively. $\mu$ here is birth/death rate (death rates were omitted in the diagram to keep it visually simple). We penalized all parameters in this model. | |
| | 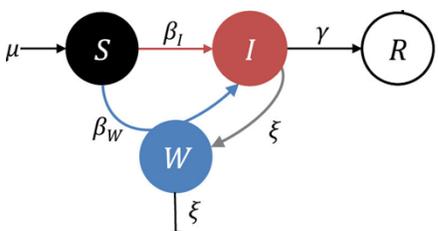 | 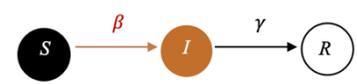 |
| | Asymptomatic model[7]: | Exponential model[7]: |
| | $S' = \mu - \beta_{IA}SI_A - \beta_{IS}SI_S - \beta_W SW - \mu S$ $\quad + \alpha_A R_A + \alpha_S R_S$ | $S' = \mu - \beta_I SI - \beta_W SW - \mu S + \alpha R$ |
| | $I_S' = q(\beta_W SW + \beta_{IS}SI_S + \beta_{IA}SI_A) - \gamma I_S$ $\quad - \mu I_S$ | $I' = \beta_I SI + \beta_W SW - \gamma I - \mu I$ $W' = \xi(I - W)$ $R' = \gamma I - \mu R - \alpha R$ |
| | $I_A' = (1-q)(\beta_W SW + \beta_{IS}SI_S + \beta_{IA}SI_A)$ $\quad - \gamma I_A - \mu I_A$ | |
| | $W' = \xi(I_A + I_S - W)$ | Where $S$, I, and $R$ represent the fractions of the population that are susceptible, infectious, and recovered, respectively. The $W$ variable is proportional to the concentration of waterborne pathogen in the environment. |
| | $R_S' = \gamma I_S - \mu I_S - \alpha_S R_S$ | |
| | $R_A' = \gamma I_A - \mu I_A - \alpha_A R_A$ | |
| | | |
| | $\tilde{y} = I_S$ | |
| | | |
| | Where $I_S$, $I_A$ represent the fractions of the population that are infectious from symptomatic and asymptomatic pathway. $R_S$, $R_A$ represent the fractions of | |

the population that are recovered from symptomatic and asymptomatic pathway. $W$ is proportional to the concentration of waterborne pathogen in the environment. Parameter $q$ is the proportion of symptomatic infection. $\tilde{y}$ represents measured output. $\alpha_S$, $\alpha_A$ are the rate that recovered individuals become susceptible after losing their immunity from symptomatic and asymptomatic pathway. Similarly, $\beta_{IS}$, $\beta_{IA}$ are transmission rate from symptomatic and asymptomatic pathway. We only penalized $\beta_{IA}$, $\beta_{IS}$, $\alpha_S$, $\alpha_A$, $q$, supposing that the other parameters were known to exist but only the symptomatic/asymptomatic infection process was uncertain.

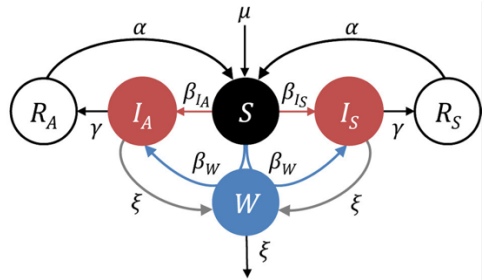

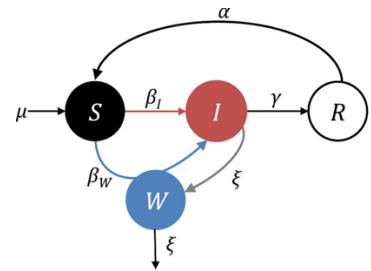

$SI_2R$ model

Variables and parameters are as described for the $SI_2R$ model above. For this model, we also included the initial conditions as unknown parameters to be estimated (as this would affect the model's ability to simplify/eliminate a transmission route). Note that this model can be shown to be unidentifiable due to the redundancy in the model transmission pathways.

$SI_2R$ model

Identical Scenario

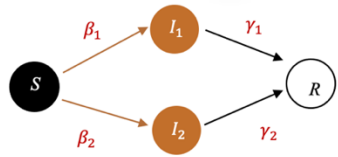

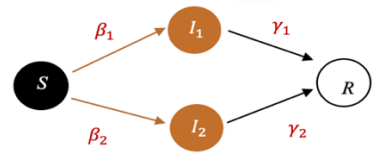

| | | |
|---|---|---|
| | *SIWR* model[7]:<br><br>Variables and parameters are as described above. We penalized all parameters in this model.<br><br>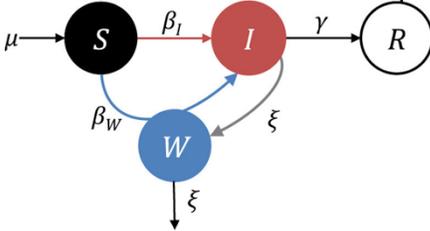 | *SIWR* model:<br><br>Variables and parameters are as described above.<br><br>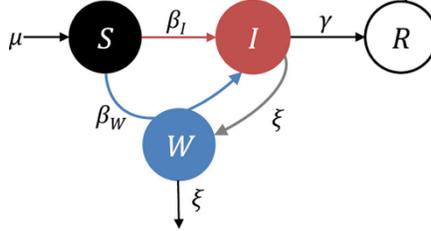 |
| Parallel Scenario | Asymptomatic + Exponential model:<br><br>$$S_0' = \mu - \beta_{IA}S_0I_A - \beta_{IS}S_0I_S - \beta_W S_0 W - \mu S_0 + \alpha_A R_A + \alpha_S R_S$$<br>$$I_S' = q(\beta_W S_0 W + \beta_{IS}S_0I_S + \beta_{IA}S_0I_A) - \gamma_a I_S - \mu I_S$$<br>$$I_A' = (1-q)(\beta_W S_0 W + \beta_{IS}S_0I_S + \beta_{IA}S_0I_A) - \gamma_a I_A - \mu I_A$$<br>$$W' = \xi_a(I_A + I_S - W)$$<br>$$R_S' = \gamma_S I_S - \mu I_S - \alpha_S R_S$$<br>$$R_A' = \gamma_a I_A - \mu I_A - \alpha_A R_A$$<br><br>$$S' = \mu - \beta_I SI - \beta_{w_1}SW - \mu S + \alpha R$$<br>$$I' = \beta_I SI + \beta_{w_1}SW - \gamma I - \mu I$$<br>$$W' = \xi(I - W)$$<br>$$R' = \gamma I - \mu R - \alpha R$$<br><br>$$\tilde{y} = w_1 I + w_2 I_S$$<br><br>Note that in this case, the alternative models are considered in parallel rather than being embedded or identical, with the measurement equation $\tilde{y}$ representing the sum of the weighted outputs of the two candidate models. This sum will allow the LASSO objective function to potentially penalize one of the models fully to zero, enabling the algorithm to perform model selection as well as parameter estimation. We penalized $\beta_{IA}$, $\beta_{IS}$, $\alpha_S$, $\alpha_A$, $\beta_W$, $\beta_I$, $\alpha$. | Exponential model |

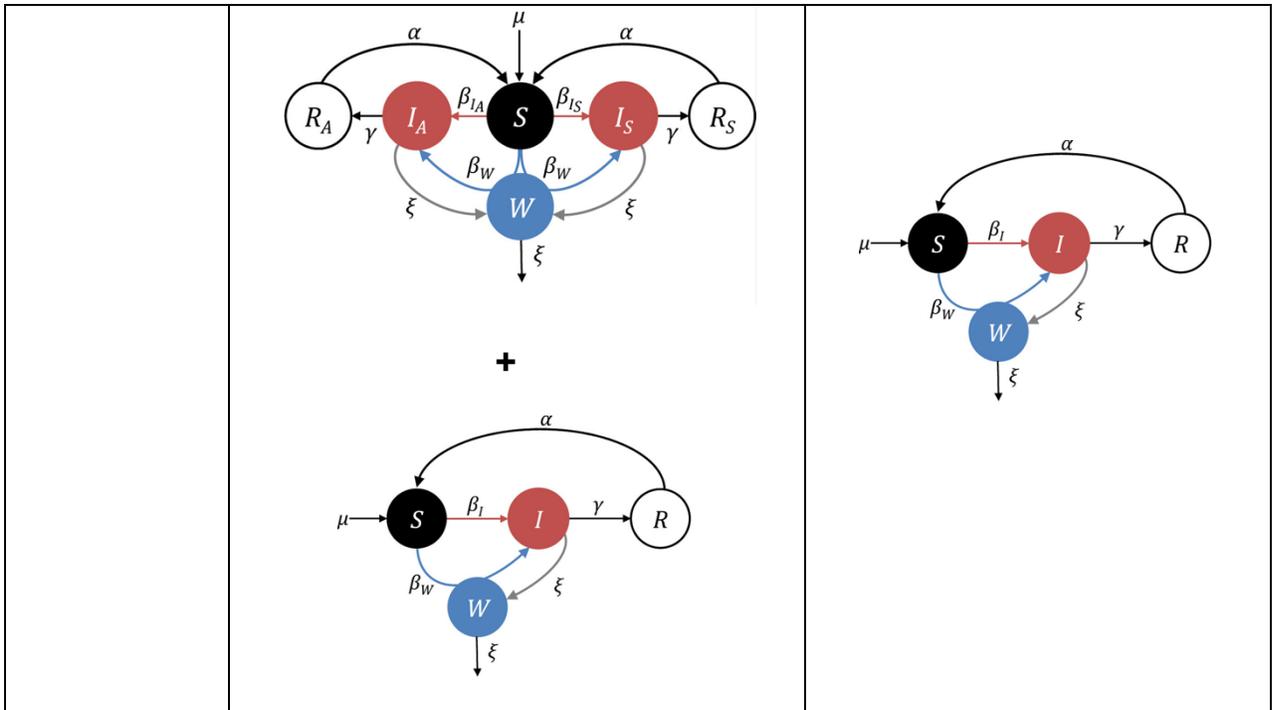

Table 1. Predictive model for training and the data that has been generated from the corresponding true model.

## 2.3 Framework application with real-world data

After completing the LASSO-ODE evaluation with simulated data, we needed to ensure that this framework is applicable to real-world data. To achieve this, we used cholera time series data collected during a major epidemic in Luanda, Angola, in 2006. This dataset includes weekly cholera case counts over a period of 168 days following the outbreak's initiation. Cholera is primarily transmitted through the ingestion of water or food contaminated with the bacterium *Vibrio cholerae*.

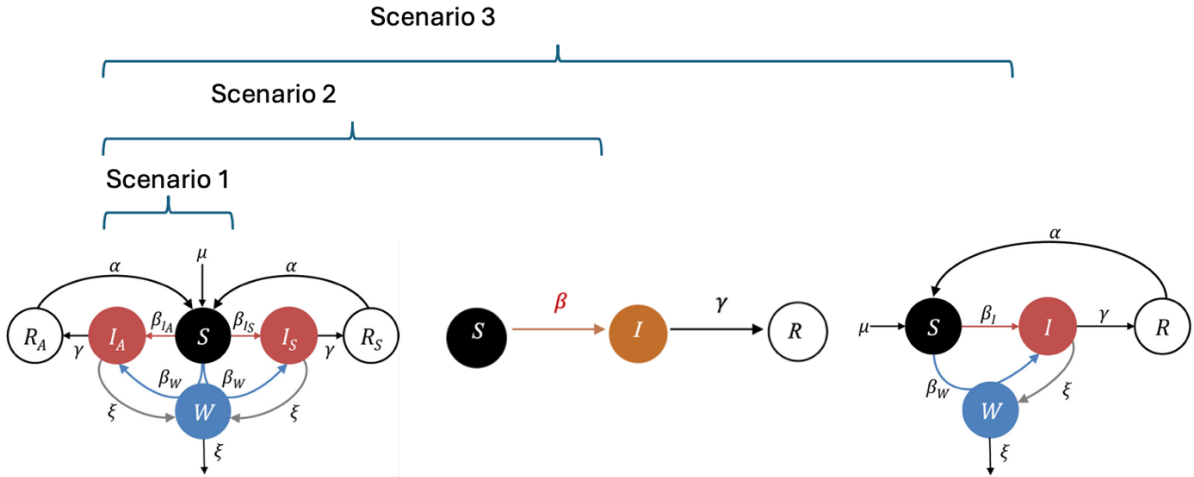

Figure 2. A schematic diagram illustrating three scenarios: 1) the Asymptomatic model alone, 2) the Asymptomatic model combined with the SIR model, and 3) the Asymptomatic model integrated with both the SIR model and the Exponential model.

We apply three predictive modeling scenarios to this data, using model drawn primarily from [7]. As shown in Figure 2, these scenarios are: 1) Asymptomatic model, 2) Asymptomatic model combined with the SIR model, and 3) Asymptomatic model combined with both the SIR model and the Exponential model (as named in Lee et al.[7]—an SIWR model with exponential loss of immunity). In the first scenario, we aim to determine if our LASSO-ODE method can identify the primary disease transmission mechanisms by highlighting the water-based parameter $\beta_W$ as the main source. In the second scenario, we introduce the SIR model, which is irrelevant, to see if our framework can correctly select the Asymptomatic model while discarding the SIR model. In the third scenario, we evaluate whether the framework can recognize two potential models—since the Exponential model is embedded in the Asymptomatic model, both models could be considered viable options.

# 3 Results

## 3.1 Scenarios from the simulated data

We evaluated all four algorithms alongside the conventional blocking cross-validation method. Overall, all four algorithms outperformed the traditional approach. Notably, Algorithm 1 excelled in cross-validation across most scenarios and was deemed the most intuitive, so the results presented below focus on Algorithm 1 (which is itself a simplified blocking cross-validation as well). However, Algorithms 3 and 4 also showed strong performance, while Algorithm 2 demonstrated mixed results. The outcomes for these methods are detailed in the Supplement. Below we detail the results for each scenario, revisiting our initial guiding questions from the Introduction:

**Embedded Scenario (addresses Q1, Q3 and Q5)**
Given that the true model is embedded in the predictive model, it is anticipated that certain estimated parameters in the predictive model will approach zero using our framework. Additionally the estimated parameters should closely align with the actual parameters used in the true model. Figure 3 illustrates that the optimal estimated parameters (denoted by red dots) selected by our framework are generally clustered near the true parameters (denoted by black dots), indicating that the LASSO-ODE framework is selecting the correct model and estimating the parameters reasonably closely to the true values. The exception is $\beta_1$ in the first scenario, which still clusters near one of the true parameter sets but shows more variability. Furthermore, all optimal estimates point toward the necessity of employing a LASSO penalty ($\lambda \neq 0$), thereby confirming the utility of the LASSO penalty in model simplification and the prevention of overfitting. Conversely, in the absence of a LASSO penalty, the estimated parameters, indicated by blue dots, display a linear distribution and deviate significantly from the actual parameters (largely driven by the lack of structural identifiability in the larger system). This contrast highlights that incorporating a LASSO penalty into the optimization function is crucial for enhancing the identifiability of our predictive model.

Moreover, all the parameters that have been selected by our framework would be interpretable and important to public health practitioner and policy makers if these examples were encountered in a real world setting: 1) When using an $SI_2R$ framework to fit the $SIR$ model, the results suggest the presence of a singular transmission route, aligning with the SIR model's structure. 2) Fitting an $SIWR$ model to $SIR$ data indicates that the disease is not transmitted via water, as the parameter $\beta_W$, representing the waterborne transmission rate, approaches zero. 3) Utilizing an *Asymptomatic* component to estimate with data from an *Exponential* model illustrates the absence of heterogeneity between asymptomatic and symptomatic transmission, evidenced by the parameters $\beta_{IA}$, representing the transmission rate from asymptomatic individuals, and $\alpha_A$, indicative of rate that recovered individuals become susceptible after losing their immunity from symptomatic pathway, both converging to zero. Overall, this initial simulated exploration of the Embedded scenario is largely successful in that the selected model is appears to be identifiable and has parameter estimates near the true values.

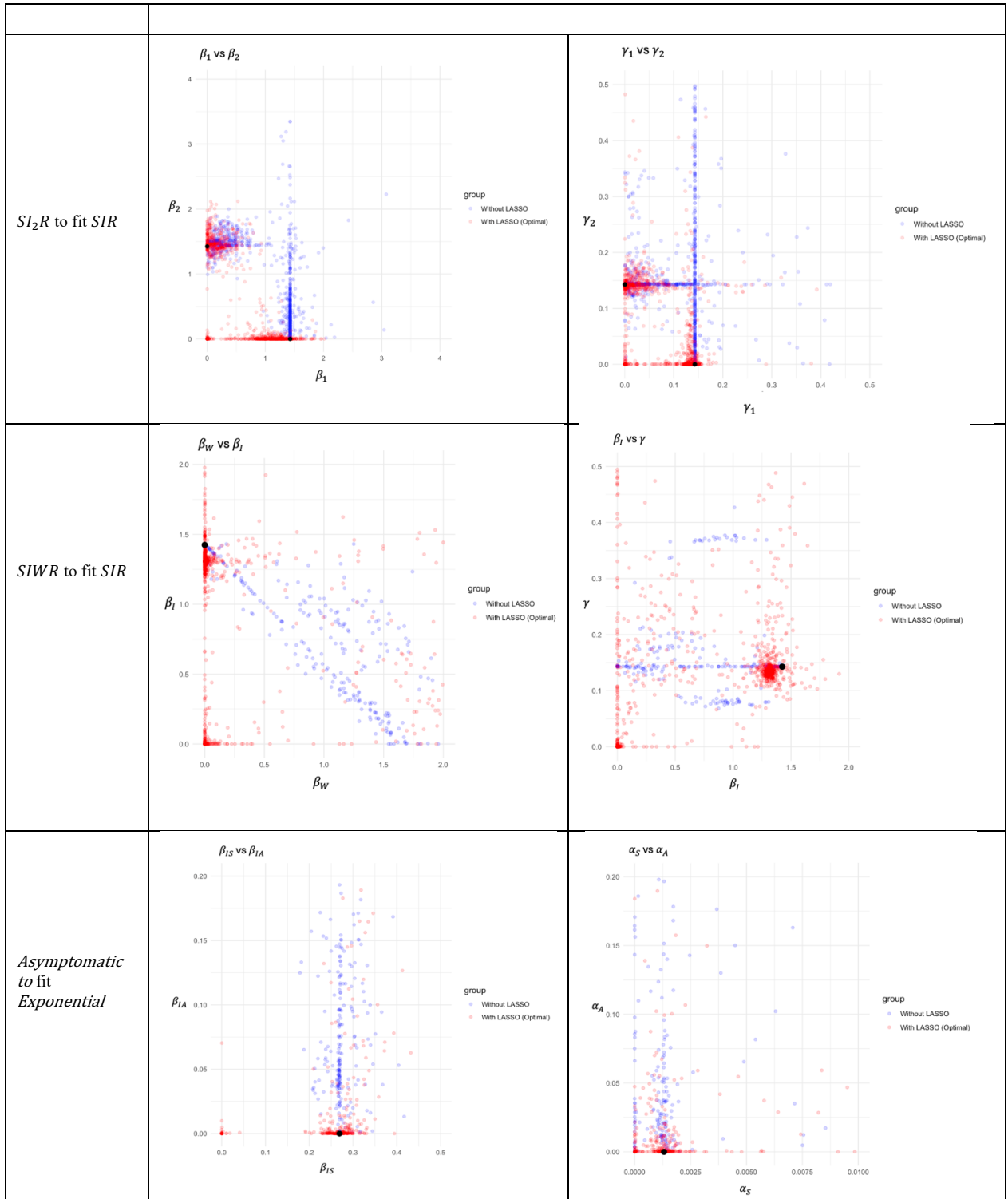

Figure 3. Example parameter estimates from embedded scenarios. In all figures, the black dot indicates the true model parameters. Each red dot corresponds to the parameter estimates obtained using a LASSO penalty in each simulation. Similarly, each blue dot represents the parameter estimates obtained without the LASSO penalty in each simulation. We have repeated this estimation 500 times because our cross-validation algorithm requires random partitioning of the data into training and testing sets.

**Identical Scenario (addresses Q2, Q3 and Q5)**

In the case where the true model and the predictive model are identical to each other, it is anticipated that no parameters will converge to zero, unless the true model is itself unidentifiable, in which case there may still be some simplification. In the case of fitting the SIWR model to itself (Figure 4), our algorithms successfully identified $\lambda = 0$ (without LASSO penalty) as the optimal parameter (see supplementary materials). This is logical, given that none of the parameters are redundant or superfluous. Essentially, this situation attests to the fact that when there is no requirement for model simplification to avoid underfitting, nor is the model overly complex for the data (e.g. unidentifiable), our methodology consistently maintains model complexity by preserving all critical parameters. Moreover, all the parameters that have been selected by our framework are interpretable: Each parameter in the predictive model, such as $\beta_W$, $\beta_I$, and others, holds distinct biological significance.

However, when fitting the SI$_2$R model to itself, our framework suggests simplifying the model ($\lambda >$ 0) because the simplified model exhibits a similar mean squared error (MSE) to the non-simplified model. This illustrates that if the true model is structurally non-identifiable (such as the SI$_2$R model), our framework may fail to select the true model because it can do equally well fitting the data with something simpler. Similarly, we would expect the SIWR model when fit to itself to eliminate the waterborne pathway when the decay rate of pathogens in the water is fast, as it is known that the model becomes practically unidentifiable and simplifies to the SIR model in this case[14,21]. When the true model is itself unidentifiable/overparameterized, the LASSO approach will simplify the system fully to the most parsimonious form matching the data—but this may actually be a desirable feature as the LASSO approach does find an identifiable submodel in this case.

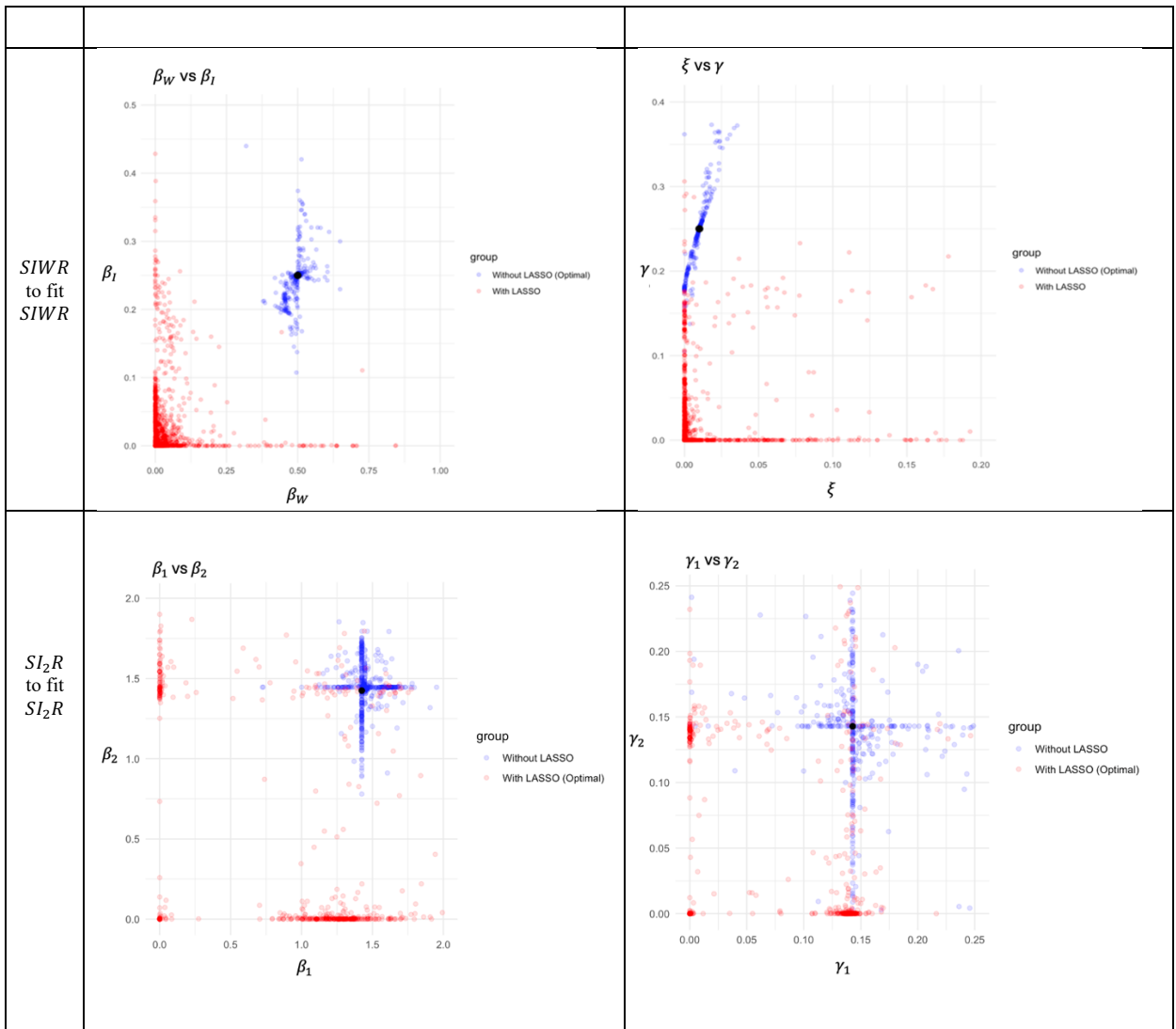

Figure 4. Model estimates from the identical scenario. In all figures, the black dot indicates the true model parameters. Each red dot corresponds to the parameter estimates obtained using a LASSO penalty in each simulation. Similarly, each blue dot represents the parameter estimates obtained without the LASSO penalty in each simulation. We have repeated this estimation 500 times because our algorithm requires random partitioning of the data into training and testing sets.

## Parallel Scenario (addresses Q4 and Q5)

In this scenario, our aim is to assess whether our framework is capable of accurately identifying the appropriate model between multiple candidate options. In this simulated case, the data set we are analyzing was generated by the exponential model. Given this, we consider both the asymptomatic and exponential models as potential fits for the data. We concurrently input these two models into our predictive function to evaluate their performance (and sum their outputs to enable LASSO to select between them). Specifically, we anticipate that the parameters of the asymptomatic model will converge towards zero, indicating that this model is unnecessary to fit the data (i.e. overfitting). Conversely, we expect the parameters of the exponential model to closely align with the original parameters used to generate the data, thereby confirming its suitability as the correct model for our data.

Figure 5 visualizes this outcome: it illustrates the asymptomatic model parameters approaching zero after the introduction of a LASSO penalty, whereas the exponential model parameters remain consistent with the true values, as indicated by the red dashed line. In addition, the mean weight parameter is 0.1 in the asymptomatic model and 0.9 in the exponential model, further suggesting that the exponential model may be the true underlying model. We note that many of the asymptomatic parameters are small even without the LASSO penalty, but are greatly reduced (in many cases explicitly to zero) when the LASSO penalty is included. Furthermore, the use of our framework appears to yield a reduction in the variance of some estimated parameters, enhancing the precision and reliability of the model selection process. Also note that the true value of alpha for the exponential model is non-zero (0.001314 before normalization), however, this parameter is highly uncertain, leading to a wide range of outlier values that expand the normalized y-axis. This is likely due to the fact that the time series is not long enough to identifiably estimate alpha, even with the LASSO penalty (particularly given that the later data is primarily used for testing rather than training).

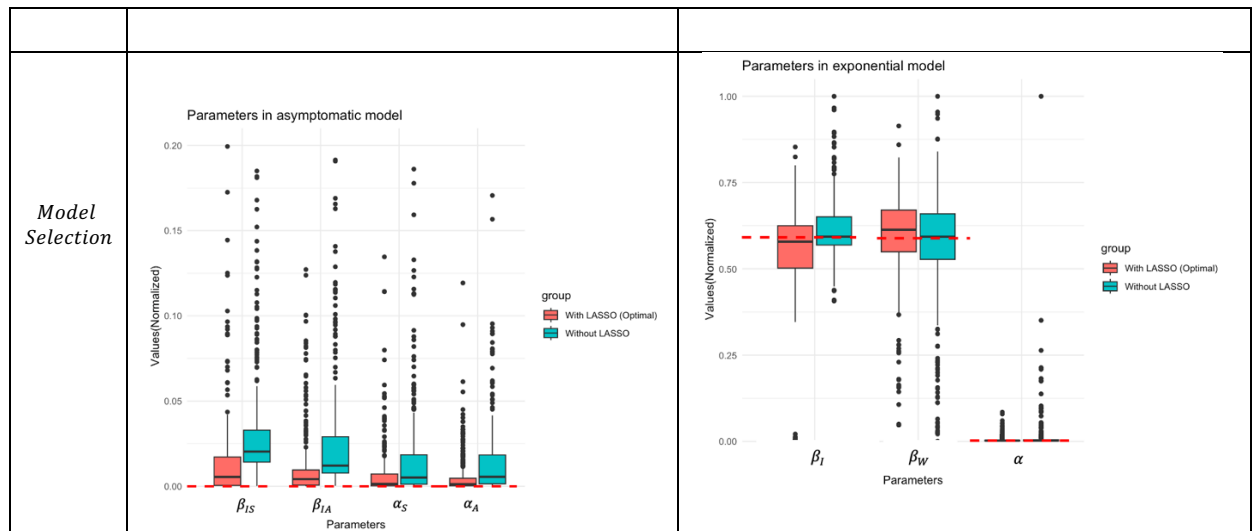

Figure 5. Model estimates from 500 simulations for parallel scenario. Red dashed line indicates the true model parameters. We applied the min-max normalization algorithm to scale all parameters uniformly, making it easier to present them in a single plot. Note that outliers (occasional poor fits) extend the y-axis of the leftmost plot but the LASSO interquartile ranges and medians are substantially closer to zero than the non-LASSO versions.

## 3.2 Scenarios from 2006 cholera outbreak in Angola

We next apply our framework to analyze real-world data obtained from the 2006 cholera outbreak in Angola.[14] For our predictive model, we apply three predictive modeling scenarios to this data: 1) the Asymptomatic model, 2) the Asymptomatic model combined with the SIR model, and 3) the Asymptomatic model combined with both the SIR model and the Exponential model. As we motioned in the Section 2.3, we would like to evaluate whether our framework can accurately capture the dynamics of cholera transmission, select the right models, and provide precise parameter estimates.

Figure 6 presents the parameter estimates from the three scenarios. In Scenario 1, it becomes evident that the primary transmission route is through contaminated water, indicated by the median transmission rate for waterborne spread ($\beta_W$) being 3.0—however the person-to-person transmission rates are not reduced to zero, suggesting that they play a some role as well. Symptomatic infections are associated with a higher transmission rate than asymptomatic ones. Additionally, our analysis reveals that the rate at which individuals with acquired immunity from symptomatic infection revert to susceptibility is much less than that for those with immunity derived from the asymptomatic pathway, which suggests that immunity following symptomatic infection is more enduring.

In Scenario 2, the framework successfully eliminates the irrelevant SIR model, as indicated by the transmission ($\beta$) and recovery ($\gamma$) rates approaching zero. It also maintains accurate estimates within the Asymptomatic model.

In Scenario 3, the framework also maintains accurate estimates for the Asymptomatic model. We note that it might appear that the Exponential model is a second viable option because $\beta_w$ in the Exponential model does not approach zero, however the weight for this model is only 0.01, compared to a median of 1.00 for the Asymptomatic model, indicating that the LASSO algorithm is continuing to choose the Asymptomatic model. This illustrates an interesting aspect of the LASSO method—the weights may be near-zero even if not all the model parameters are fully reduced by the penalization. This may indicate a need for a more finely resolved estimation of λ. The SIR model is again accurately disregarded.

Table 2 presents the parameter estimates from the asymptomatic model across three scenarios. In all three scenarios, contaminated water is identified as the main transmission source, as indicated by the comparatively higher value of $\beta_W$ compared to person-to-person transmission. Additionally, the loss of immunity rates between asymptomatic and symptomatic individuals differ significantly, underscoring the need to separate these two compartments. Overall, the similarity in parameter estimates across the scenarios underscores the robustness of the LASSO-ODE method.

| Predictive model | Estimated parameters in the predictive model |
|---|---|
| Scenario 1 | 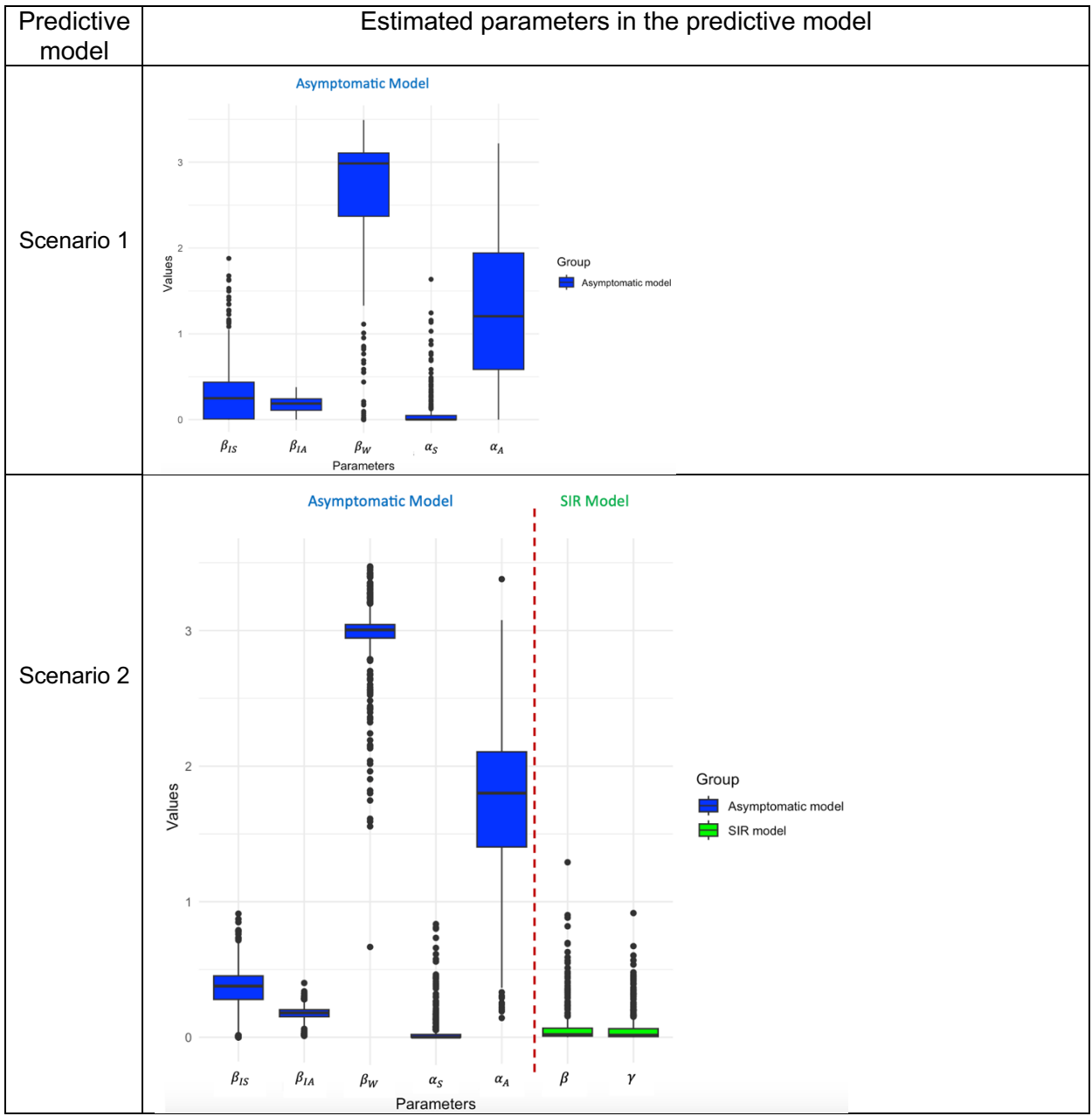 |
| Scenario 2 | |

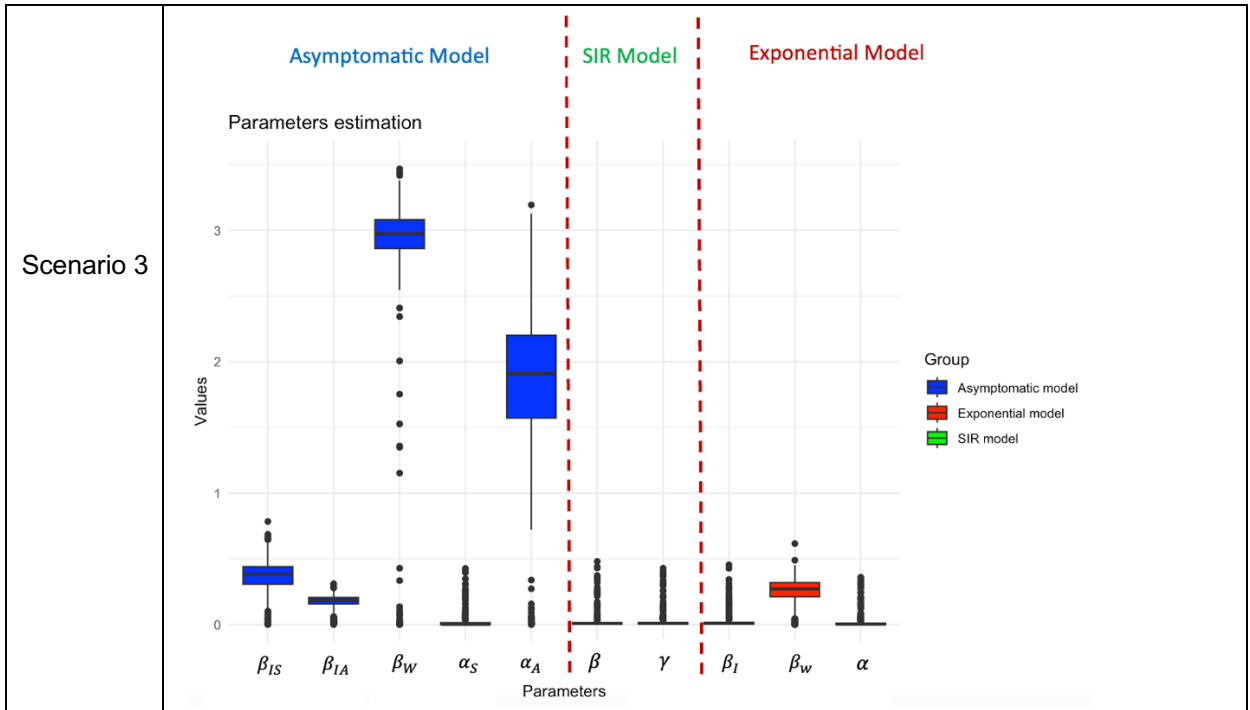

Figure 6. Model estimates from 500 times simulations for three distinct scenarios using data from 2006 cholera outbreak in Angola.

| Parameters in the Asymptomatic model | Value (median [1st quantile, 3rd quantile]) | | |
|---|---|---|---|
| | Scenario 1 | Scenario 2 | Scenario 3 |
| $\beta_S$ | 0.25 [0.01, 0.44] | 0.38 [0.28, 0. 45] | 0.38 [0.31, 0.44] |
| $\beta_A$ | 0.18 [0.11, 0.24] | 0.18 [0.15, 0.20] | 0.18 [0.16, 0.21] |
| $\beta_W$ | 3.01 [2.95, 3.48] | 3.01 [2.96, 3.06] | 3.00 [2.86, 3.08] |
| $\alpha_S$ | 0.0012 [0.0001, 0.049] | 0.0012 [0.00007, 0.02] | 0.0010 [0.00004, 0.01] |
| $\alpha_A$ | 1.20 [0.58, 1.94] | 1.80 [1.40, 2.11] | 1.91 [1.57, 2.20] |
| $k^{-1}$ | 0.000010 [0.000010, 0.000011] | 0.000010 [0.000010, 0.000010] | 0.000010 [0.000010, 0.000010] |
| $\omega_1$ | - | 1.00 [0.99, 1.00] | 1.00 [1.00, 1.00] |
| $\omega_2$ | - | 0.00 [0.00, 0.01] | 0.00 [0.00, 0.00] |
| $\omega_3$ | - | - | 0.01 [0.002, 0.05] |

Table 2. The parameters estimation in asymptomatic model using 2006 cholera data in Angola. Note: $\omega_1, \omega_2, \omega_3$ here represent the weight from asymptomatic model, SIR model, and exponential model.

Figure 7 below illustrates how the model in different scenarios fit our data. Overall, the data aligns well with the prediction line, which represents the median of the predictions. However, the confidence interval widens after day 80, with the confidence region extending away from the data. This is likely in part due to the real-world data being influenced by several non-model-related factors, such as reporting lapses and other issues, but also likely because the later data is more often used for testing rather than training and so is more likely to show worse performance, illustrating one potential drawback of parameter estimation methods such as LASSO that require cross-validation. Some of our other proposed algorithms for cross validation do not require the use of later data for testing (Algorithms 3 and 4), which could be a potential option to help mitigate this issue.

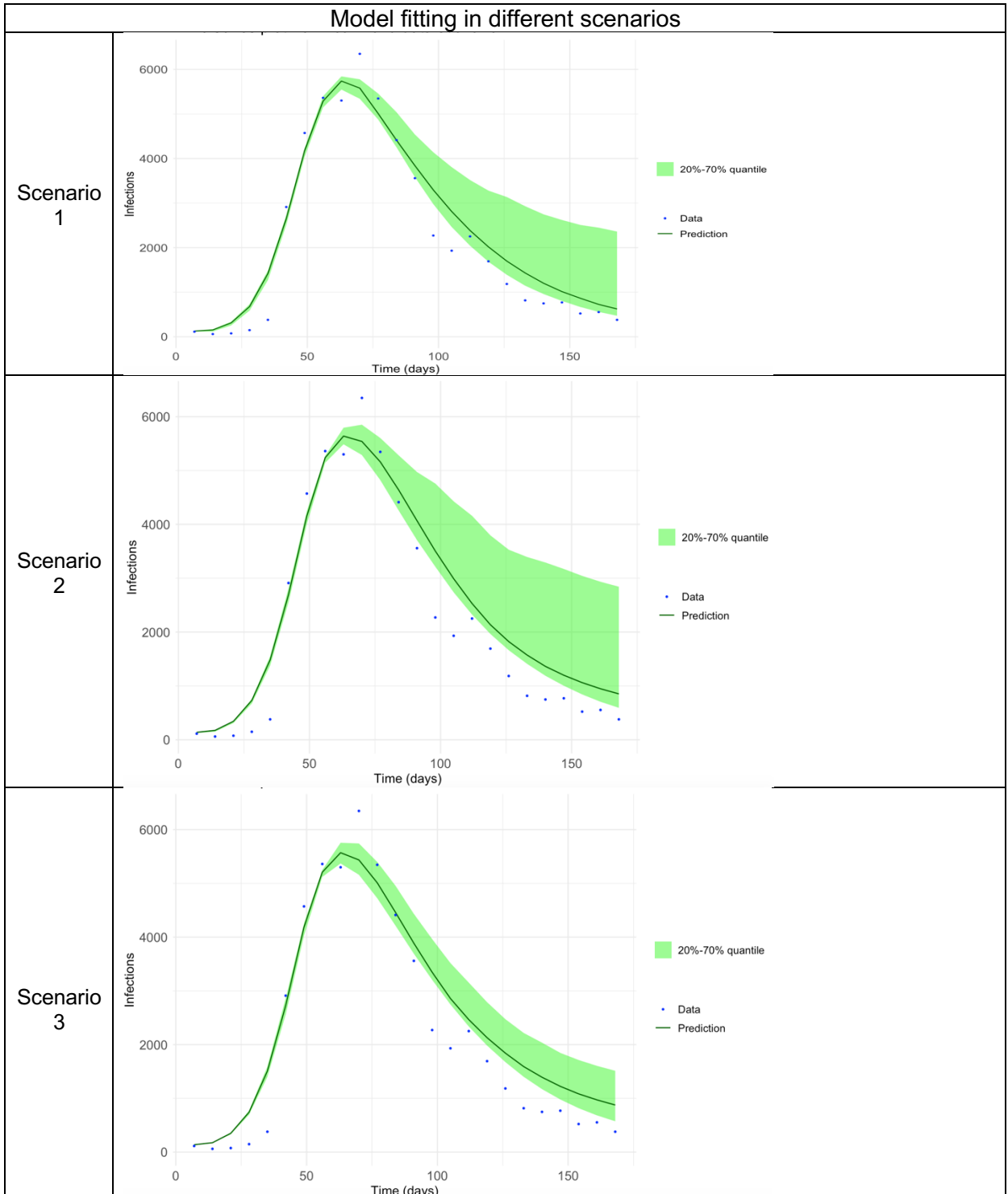

Figure 7. Model best fit to cholera data from the 2006 Angola epidemic. Note: We present the predictions from the 20th to the 70th percentile because this range provides narrower confidence bounds, indicating more robust estimations

# 4 Discussion

In this work, we evaluated several alternative model structure scenarios (embedded, identical, parallel) to explore how a LASSO penalty term could improve the linking of mechanistic models with data, by simultaneously evaluating model selection, dimension reduction/identifiability, and parameter estimation. In our explorations, we observed several broader themes, described below.

*Model identifiability can be addressed real-time during parameter estimation, and depends on our objective function.* Mechanistic models often have issues with model identifiability, as illustrated by the linear trends representing identifiable combination relationships observed in our non-LASSO model fits. By modifying our evaluation function to include a LASSO penalty, we found that the model fit was preserved but models were consistently reduced to an identifiable form. This illustrates the known point that model identifiability is not absolute; rather, it is influenced by several factors, including the choice of evaluation/objective function. However, we want to note that there remains "no free lunch"—there is still no way to reliably estimate all the original model parameters from the data at hand, but at least this approach provides a way to explore the space of identifiable reduced models in an automatic way, while generating parameter estimates as well. A key contribution of this paper is demonstrating that penalized regression hyperparameters can enable us to address unidentiability issues "on-the-fly" as part of the parameter estimation process, without necessarily requiring additional identifiability analysis. However, it is worth noting that identifiability analysis would still be needed to understand the structures discovered via LASSO fitting.

*Clustered parameter estimates indicate identifiability.* Our observations from multiple scenarios demonstrate that when the LASSO penalty is included, optimal parameter estimates consistently cluster near the actual values after several numerical simulations. This behavior indicates model identifiability, characterized by parameter estimates that form a condensed cluster. In contrast, an unidentifiable model produces parameter estimates that scatter broadly, resembling a uniform dense line or curve. We note that this clustering pattern is a useful indicator of identifiability even when the uncertainty/confidence ranges for the parameters are still large. However, it is also important to note that we only tested relatively simple identifiable combinations in this study, but larger/more complex identifiable combinations involving more degrees of freedom may result in more complex patterns and should be explored in future work. Nonetheless, the results showcased in this study confirm that our proposed framework reliably converges on true parameters, resulting in a distinct clustering pattern that signifies identifiable parameter estimates.

*Model selection can also be integrated real-time into the parameter estimation and identifiability process.* Doing so enables us to simultaneously evaluate the two major sources of uncertainty when linking models and data: parameter uncertainty and model structure uncertainty (sometimes called distinguishability). We tested two approaches to examining model structure uncertainty in this study—reduction of a larger, potentially unidentifiable model to a reduced submodel (which highlights the strong link between model and parameter uncertainty), and selection between multiple parallel models via summing their outputs (so that the LASSO penalty can select between them). Both approaches worked well in our initial tests here, and realistically probably many real-world applications will involve some degree of both the embedded and parallel approaches, as many models that we may test in parallel are likely to be unidentifiable. The LASSO approach explored here allows the natural interaction between parameter identifiability and model structure uncertainty to be explored explicitly, rather than the model selection being split into a separate process that occurs after parameter estimation (e.g. via the Akaike information criterion (AIC) and similar approaches). This separation of parameter and model structure uncertainty can lead to issues as unidentifiable models can be penalized by the AIC when their actual effective degrees of freedom are substantially smaller than their apparent parameters due to identifiability issues (e.g. as illustrated in Cortez et. al[18]). Combining the two enables the identifiability and model selection processes to be connected and considered simultaneously.

*LASSO-ODE and SINDy[25] address fundamentally different but related research questions.* As mentioned earlier, LASSO-ODE is a framework for selecting the most suitable and identifiable model from existing candidates, while SINDy constructs ODE models de novo without any prior information. In practical scenarios when using epidemiological data, we often only have access to highly noisy, incomplete, and uncertain data on the number of infected individuals. Consequently, LASSO-ODE may be sometimes more realistic to apply even if it has more restricted applicability to an established set of models, as it does not require comprehensive, clean data from every compartment of an ODE system to estimate parameters. In contrast, SINDy requires a complete and ideally relatively non-noisy dataset to operate successfully (although research on these areas is currently active [add some of the more recent SINDY-related papers]). Additionally, the LASSO/penalized regression approach is useful beyond questions of model discovert—for example, LASSO approaches can also be used to test the structural identifiability of an existing model. For example, if we want to determine whether a smaller model is identifiable, we can fit several larger models to the data that has been simulated from this smaller model. If these larger models eventually shrink to the smaller model, it suggests that the smaller model is structurally identifiable.

*One-fold cross-validation is sufficient for data with a single epidemic peak.* All four of our algorithms utilize only one-fold cross-validation. Although we experimented with k-fold (k≥2) cross-validation, as is common with traditional blocking methods, the resulting estimates were poor (see supplementary S1). This occurs because the most critical information (the data that include the peak of the epidemic curve) for parameter estimation is typically concentrated in the first fold. As a result, the remaining data (the right tail of the epidemic curve) in subsequent folds can lead to inaccurate estimates of the underlying mechanisms.

While our framework demonstrates the ability of LASSO/penalized estimation methods to resolve model identifiability and selection, it does have several limitations:

*LASSO-ODE framework can fail to detect the ground truth parameters when the true model is structurally unidentifiable from the data available.* Our objective is to ascertain parameters that precisely fit the data, yet the structural unidentifiability of our model may lead to a range of distinct optimized solutions for different identifiable reductions of the true model. Consider the case where the $SI_2R$ model is deployed for self-fitting (i.e., the identical scenario); in many cases we are likely to converge on a simplified version, e.g. the SIR model, as a result of $SI_2R$ 's inherent structural unidentifiability, which stems from a high level of symmetry in its design. This is in some sense a weakness of our approach—it cannot correctly determine the underlying model in cases where the true model was unidentifiable. However, this is also a situation where one likely would not want to work with the true model unless more data could be collected—the "true" model in this situation is more complex than the data warrants.

*Not all algorithms for cross-validation (CV) are efficient.* As shown in the supplementary materials, not all algorithms provide accurate estimates. For instance, in Algorithm 2, we divide the data around the peak area, splitting the most crucial information into two pieces, which can possibly lead to the incorrect selection of λ. Further, Algorithm 3 suffers from a data leakage issues since the testing data is generated from the training data. Despite this, Algorithm 3 can still be useful for sensitivity analysis. However, compared to traditional blocking cross-validation, all four algorithms performed substantially better, as demonstrated in the supplementary materials. Furthermore, these four algorithms can be applied to various types of data not considered here, such as multi-wave epidemics or non-infectious disease data like cancer research. The performance of these algorithms may differ in these cases given the potentially more complex temporal dynamics.

*LASSO with cross validation can be computationally expensive.* Our framework requires multiple rounds of parameter estimation, and so can be computationally intensive. In this study, we tested only five λ values for each scenario, and the models were relatively simple. Even so, some scenarios took on the order of hours to reach a converged set of estimates including a cross-validated LASSO penalty λ. However, in real-world applications, more λ values may need to be tested, and models will often be more complex. Consequently, the use of supercomputers or parallel computing may be necessary to operate our framework efficiently.

In spite of these limitations, we found that integrating LASSO/penalized estimation methods into mechanistic modeling methods offers a promising approach to elucidating uncertainties. In this study, our prediction framework is rooted in traditional ordinary differential equation (ODE) models, which are both interpretable and well-suited for epidemic forecasting, scenario exploration, and other common public health use cases. However, we enhance this foundation by incorporating machine learning hyperparameters, such as the LASSO penalty in the objective function, to refine our parameter estimation and incorporate identifiability and model selection. This method illustrates an example approach to maintaining the explanatory power of the mechanistic model while benefiting from the increased predictive accuracy afforded by machine learning-inspired techniques, an idea that has shown great promise across a range of infectious disease applications.[10]

*Conclusions.* In this paper, we propose the use of the LASSO penalty for ordinary differential equations (ODEs) in the context of infectious disease modeling. Our framework demonstrates how the LASSO penalty can allow for simultaneously address model identifiability, selection, and parameter estimation, providing useful and interpretable parameter estimates that contribute to the inference of transmission processes. This aspect is beneficial to public health practitioners and policymakers, as it aids in informed decision-making and enables the generation of what-if scenarios, intervention testing, and counterfactuals. Additionally, the new cross-validation techniques for time series data introduced in our work offer potential new tools for data scientists across a range of applications.

**Supplementary**

**Supplementary Materials Section 1: Alternative algorithms for cross-validation.**

Here we provide the algorithms and brief overview for our alternative cross-validation methods.

---

**Algorithm 2: Split around peak**

---

Given that previous studies of model parameter sensitivities for SIR models have shown that crucially informative data for parameter estimation from single epidemic models is predominantly concentrated around the apex of the infectious curve.[15] we aim to position the dividing line between the training and testing datasets within this peak vicinity. This strategy ensures that both sets contain pertinent information. However, this approach only makes sense for infectious disease epidemic scenarios where there is likely to be a single peak in the dynamics.

---

1: **Input:** Observed time series infections $\mathbf{I} = \{I_t\}_{t=1}^{T}$ , where $T$ here is the total length of time.

   A range of $\boldsymbol{Lambda} = \boldsymbol{\lambda}$.

   Initial guess of parameters in the prediction model $\boldsymbol{\theta_0} = \{\theta_i\}_{i=1}^{P}$,

   where $P$ here is the total number of parameters in the model.

   The total number of simulation times $C$.

2. **Output:** Corresponding optimal parameters $\overline{\boldsymbol{\theta^*}}$

3. **for** $\lambda$ in $\boldsymbol{Lambda}$, **do**

4.    k = 1

5.    $t_{peak} = \underset{t}{\textbf{argmax}}\{\ \mathbf{I_t}\ \}$

6.    **while** (k < $C$), **do**

7.       Cutting_point $\sim$ U $(t_{peak} - \sigma, t_{peak} + \sigma)$, $\sigma$ here is a fixed interval

8.       $\mathbf{I_{train,k}} = \{I_t\}_{t=1}^{\text{Cutting\_point}}$

9.       $\mathbf{I_{test,k}} = \{I_t\}_{t=\text{Cutting\_point}+1}^{\text{T}}$

10.      $\boldsymbol{\theta_{k0}} = \boldsymbol{\theta_0} + \boldsymbol{\mathcal{N}}(\boldsymbol{\theta_0}, \ 0.2)$,   0.2 here can be other positive number

11.      $\boldsymbol{\theta_{k,\lambda}} = \underset{\boldsymbol{\theta}}{\textbf{argmin}}\{\ \sum(\widetilde{\mathbf{I_{train,k}}} - \mathbf{I_{train,k}})^2 + \lambda\sum_j |\ \vartheta_j\ |\ \}$ given $\boldsymbol{\theta_{k0}}$, $I_{t=1}$, and $\lambda$

            where $\widetilde{\mathbf{I_{train,k}}}$ is the measurement function,

            $\vartheta_j$ is the parameter that need to be penalized

12.      $\widetilde{\mathbf{I_{test,k}}}$ = The predicted number of infections given $\boldsymbol{\theta_{k,\lambda}}$, and $I_{t=\text{Cutting\_point}+1}$

13.      $\text{MSE}_{test,k} = \frac{1}{length(\mathbf{I_{test,k}})}\sum(\widetilde{\mathbf{I_{test,k}}} - \mathbf{I_{test,k}})^2$

14.      k = k+1

15.   **end while**

16.   $\boldsymbol{\theta_{\lambda}} = \{\boldsymbol{\theta_{k,\lambda}}\}_{k=1}^{C}$ is the collection of $\boldsymbol{\theta_{k,\lambda}}$ from each $k$th simulation under $\lambda$

17.   $\overline{\text{MSE}_{\lambda,\boldsymbol{\theta_\lambda}}}$ = mean (or median) of $\{\text{MSE}_{test,k}\}_{k=1}^{C}$ from $C$ times simulation under $\lambda$

18. **end for**

19. $\lambda^* = \underset{\lambda, \boldsymbol{\theta}_\lambda}{\operatorname{\textbf{argmin}}} \{ \overline{\text{MSE}_{\lambda, \boldsymbol{\theta}_\lambda}} \}$

20. $\boldsymbol{\theta}^*$ is the $\boldsymbol{\theta}_\lambda$ under $\lambda^*$

21. $\overline{\boldsymbol{\theta}^*}$ = mean of $\boldsymbol{\theta}^*$ from $C$ times simulation under $\lambda^*$ (you can also use median of $\boldsymbol{\theta}^*$)

22. Return $\overline{\boldsymbol{\theta}^*}$ (you can also return $\boldsymbol{\theta}^*$ to see the entire distribution of $C$ times estimation)

---

**Algorithm 3: Generate different data based on random Gaussian noise**

---

Here we use a bootstrapping-inspired approach produced test data by introducing random Gaussian noise across the entire dataset in Algorithm 3 (where our Gaussian is based on the distribution of residuals in the original fit). This approach allows us to evaluate the sensitivity of the estimated parameters derived from the training data to the presence of random perturbations, thereby assessing the stability of our model. Additionally, this ensures that the test and training data exhibit similar patterns, which in turn renders the mean squared error (MSE) derived from the test data a more dependable measure of performance. However, this approach is likely to have data leakage issues and does not test different underlying dynamics, just different realizations of the noise. Therefore, one might use this algorithm only to do sensitivity analysis to test if model is stable after adding random noise.

---

1: **Input:** Observed time series infections $\mathbf{I} = \{I_t\}_{t=1}^T$ , where $T$ here is the total length of time.
   A range of $\boldsymbol{Lambda} = \boldsymbol{\lambda}$.
   Initial guess of parameters in the prediction model $\boldsymbol{\theta_0} = \{\theta_i\}_{i=1}^P$,
   where $P$ here is the total number of parameters in the model.
   The total number of simulation times $C$.

2. **Output:** Corresponding optimal parameters $\overline{\boldsymbol{\theta}^*}$

3. **Residual** = $\mathbf{I} - \bar{\mathbf{I}}$, where $\bar{\mathbf{I}}$ is the estimated infections using least square

4. $\text{Residual}_{mean}$ = mean of **Residual**

5. $\text{Residual}_{variance}$ = variance of **Residual**

6. **for** $\lambda$ in $\boldsymbol{Lambda}$, **do**

7.     k = 1

8.     **while** (k < $C$), **do**

9.         $\mathbf{I}_{train,k} = \mathbf{I}$

10.         $\mathbf{I}_{test,k} = \mathbf{I}_{train,k} + \mathcal{N}(\text{Residual}_{mean} , \text{Residual}_{variance})$

11.         $\boldsymbol{\theta}_{k0} = \boldsymbol{\theta_0} + \mathcal{N}(\boldsymbol{\theta_0} , 0.2)$, 0.2 here can be other positive number

12.         $\boldsymbol{\theta}_{k,\lambda} = \underset{\boldsymbol{\theta}}{\operatorname{\textbf{argmin}}} \{ \sum (\widetilde{\mathbf{I}_{train,k}} - \mathbf{I}_{train,k})^2 + \lambda \sum_j |\vartheta_j| \}$ given $\boldsymbol{\theta}_{k0}$, $I_{train,k,t=1}$, and $\lambda$
           where $\widetilde{\mathbf{I}_{train,k}}$ is the measurement function,
           $\vartheta_j$ is the parameter that need to be penalized

13.         $\widetilde{\mathbf{I}_{test,k}}$ = The predicted number of infections given $\boldsymbol{\theta}_{k,\lambda}$, and $I_{test,k,t=1}$

14.     $\mathrm{MSE}_{test,k} = \frac{1}{length(\mathbf{I}_{test,k})} \sum (\widetilde{\mathbf{I}_{test,k}} - \mathbf{I}_{test,k})^2$

15.     k = k+1

16.     **end while**

17.     $\boldsymbol{\theta}_{\lambda} = \{\boldsymbol{\theta}_{k,\lambda}\}_{k=1}^{C}$ is the collection of $\boldsymbol{\theta}_{k,\lambda}$ from each $k$th simulation under $\lambda$

18.     $\overline{\mathrm{MSE}}_{\lambda,\boldsymbol{\theta}_{\lambda}}$ = mean (or median) of $\{\mathrm{MSE}_{test,k}\}_{k=1}^{C}$ from $C$ times simulation under $\lambda$

19.  **end for**

20.  $\lambda^* = \underset{\lambda,\boldsymbol{\theta}_{\lambda}}{\mathbf{argmin}}\{\ \overline{\mathrm{MSE}}_{\lambda,\boldsymbol{\theta}_{\lambda}}\ \}$

21.  $\boldsymbol{\theta}^*$ is the $\boldsymbol{\theta}_{\lambda}$ under $\lambda^*$

22.  $\overline{\boldsymbol{\theta}^*}$ = mean of $\boldsymbol{\theta}^*$ from $C$ times simulation under $\lambda^*$ (you can also use median of $\boldsymbol{\theta}^*$)

23.  Return $\overline{\boldsymbol{\theta}^*}$  (you can also return $\boldsymbol{\theta}^*$ to see the entire distribution of $C$ times estimation)

---

**Algorithm 4: Generate different data based on initial conditions (simulated data only)**

---

Algorithm 4 is designed primarily for use with simulated data, as it requires the application of known true parameters and various initial values to create the dataset. The underlying concept is to ensure that both the training and test data are generated from the same underlying mechanism, thereby maintaining consistency in the datasets without introducing the data leakage issues encountered with the Algorithm 3. Although Algorithm 4 cannot be used for epidemic forecasting, it is useful for exploring the identifiability of model structures. For instance, if we want to assess whether a smaller model (the true model) is identifiable, we can fit several larger models (predictive models) to data simulated from this smaller model. If these larger models eventually reduce to the smaller model, it indicates that the smaller model is structurally identifiable.

---

1: **Input:** Ground truth parameters $\boldsymbol{\theta}$.

       Initial conditions to generate train dataset $\boldsymbol{K_{train\_0}}$.

       Initial conditions to generate test dataset $\boldsymbol{K_{test\_0}}$.

       True model $f(\boldsymbol{\theta}, \boldsymbol{K_{train_0}}, \boldsymbol{K_{test\_0}})$ that used to generate simulated data.

       A range of $\boldsymbol{Lambda} = \boldsymbol{\lambda}$,

       Initial guess of parameters in the prediction model $\boldsymbol{\theta_0} = \{\theta_i\}_{i=1}^{P}$,

       where $P$ here is the total number of parameters in the model.

       The total number of simulation times $C$.

2. **Output:** Corresponding optimal parameters $\overline{\boldsymbol{\theta}^*}$

3. **for** $\lambda$ in $\boldsymbol{Lambda}$, **do**

4.     k = 1

5.     $\mathbf{I}_{train} = f(\boldsymbol{\theta}, \boldsymbol{K_{train\_0}})$, where $f(x)$ here is the true model

6.     $\mathbf{I}_{test} = f(\boldsymbol{\theta}, \boldsymbol{K_{test\_0}})$, where $f(x)$ here is the true model

7.      **while** ($k < C$), **do**

8.          $\boldsymbol{\theta_{k0}} = \boldsymbol{\theta_0} + \boldsymbol{\mathcal{N}(\theta_0}, \ 0.2)$,   0.2 here can be other positive number

9.          $\boldsymbol{\theta_{k,\lambda}} = \underset{\boldsymbol{\theta}}{\textbf{argmin}} \{ \sum (\widetilde{\mathbf{I_{train,k}}} - \mathbf{I_{train}})^2 + \lambda \sum_j | \vartheta_j | \ \}$ given $\boldsymbol{\theta_{k0}}$, $I_{train,t=1}$, and $\lambda$

                where $\mathbf{I_{train,k}}$ is the measurement function,

                $\vartheta_j$ is the parameter that need to be penalized

10.         $\widetilde{\mathbf{I_{test,k}}}$ = The predicted number of infections given $\boldsymbol{\theta_{k,\lambda}}$, and $I_{test,t=1}$

11.         $\text{MSE}_{test} = \frac{1}{length(\mathbf{I_{test}})} \sum (\widetilde{\mathbf{I_{test,k}}} - \mathbf{I_{test}})^2$

12.         $k = k+1$

13.      **end while**

14.      $\boldsymbol{\theta_\lambda} = \{ \boldsymbol{\theta_{k,\lambda}} \}_{k=1}^C$ is the collection of $\boldsymbol{\theta_{k,\lambda}}$ from each $k$th simulation under $\lambda$

15.      $\overline{\text{MSE}_{\lambda,\boldsymbol{\theta_\lambda}}}$ = mean (or median) of $\{ \text{MSE}_{test,k} \}_{k=1}^C$ from $C$ times simulation under $\lambda$

16. **end for**

17. $\lambda^* = \underset{\lambda,\boldsymbol{\theta_\lambda}}{\textbf{argmin}} \{ \ \overline{\text{MSE}_{\lambda,\boldsymbol{\theta_\lambda}}} \ \}$

18. $\boldsymbol{\theta^*}$ is the $\boldsymbol{\theta_\lambda}$ under $\lambda^*$

19. $\overline{\boldsymbol{\theta^*}}$ = mean of $\boldsymbol{\theta^*}$ from $C$ times simulation under $\lambda^*$ (you can also use median of $\boldsymbol{\theta^*}$)

20. Return $\overline{\boldsymbol{\theta^*}}$ (you can also return $\boldsymbol{\theta^*}$ to see the entire distribution of $C$ times estimation)

**Supplementary Materials Section 2: Results from alternative algorithms for cross-validation.**

Tables S1 to S5 present the model estimates using Algorithms 2 to 4. Additionally, Table S1 includes results from the conventional cross-validation (traditional method) for comparison, where we employed multifold blocking cross-validation:

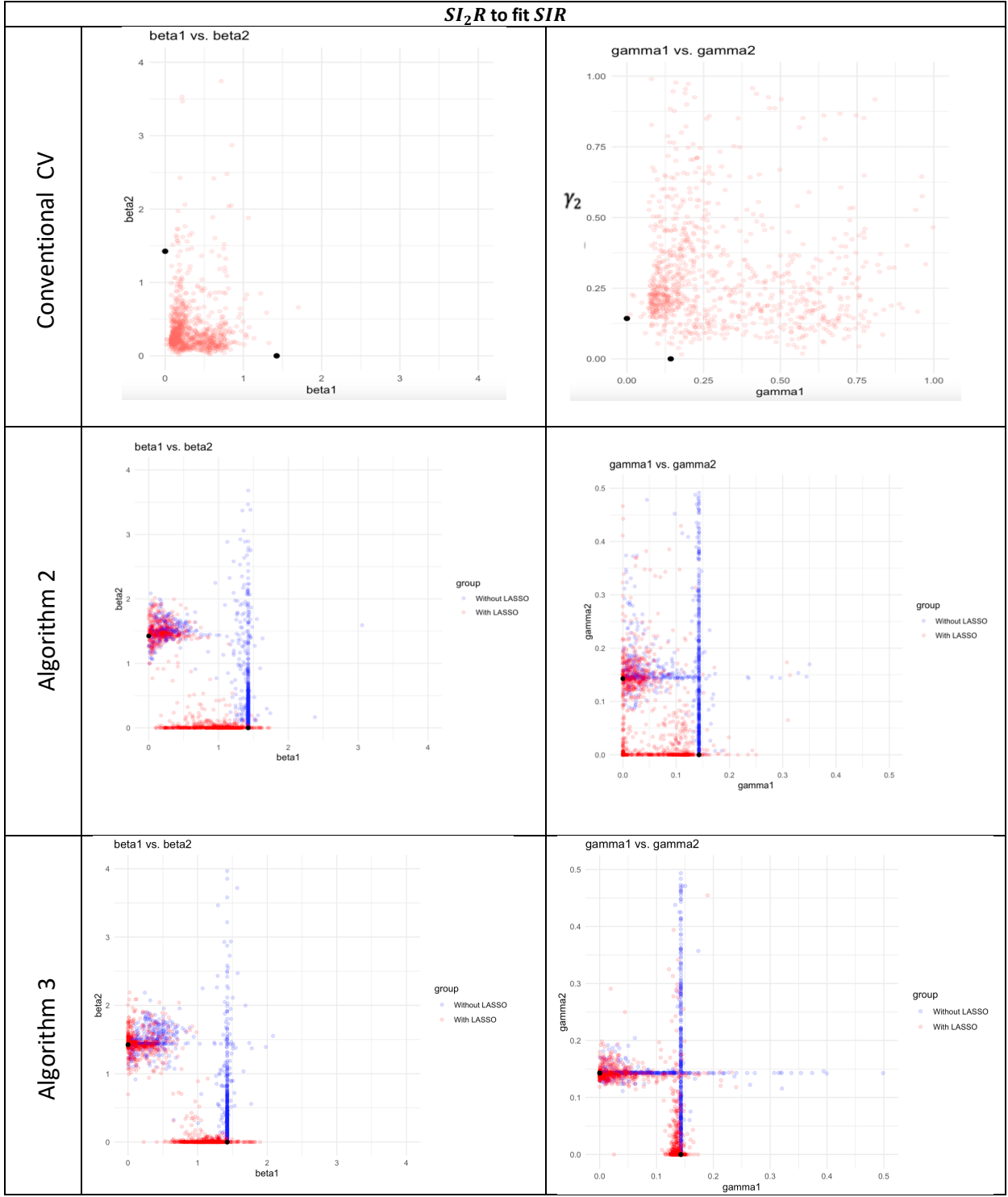

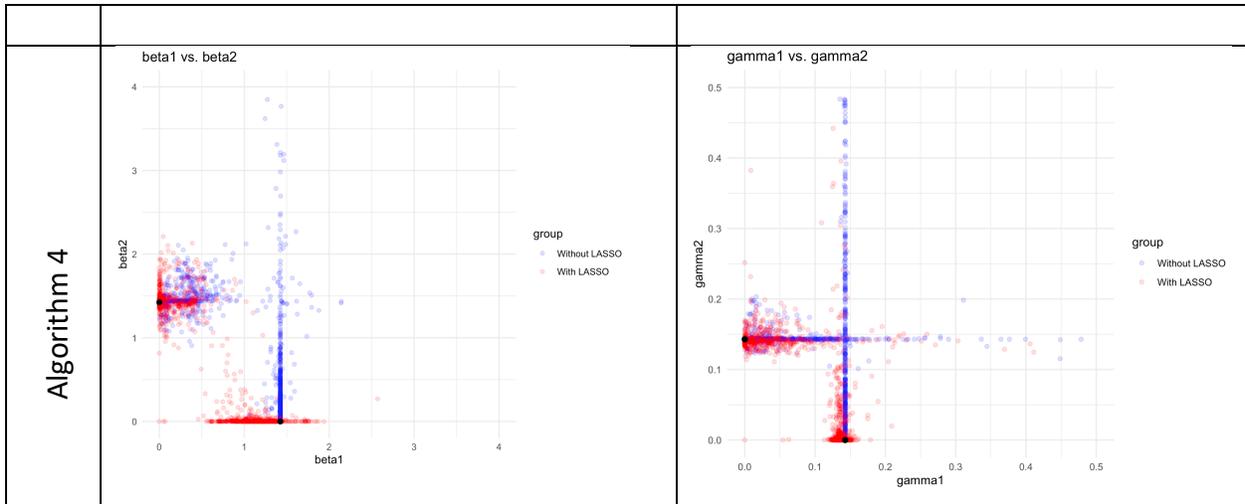

S1. Model estimates from $SI_2R$ to fit $SIR$ scenario (Algorithm 2 - 4). Note: Here beta1, beta2, gamma1, gamm2 are equivalent to $\beta_1$, $\beta_2$, $\gamma_1$, $\gamma_2$ in Table 1 in the main draft.

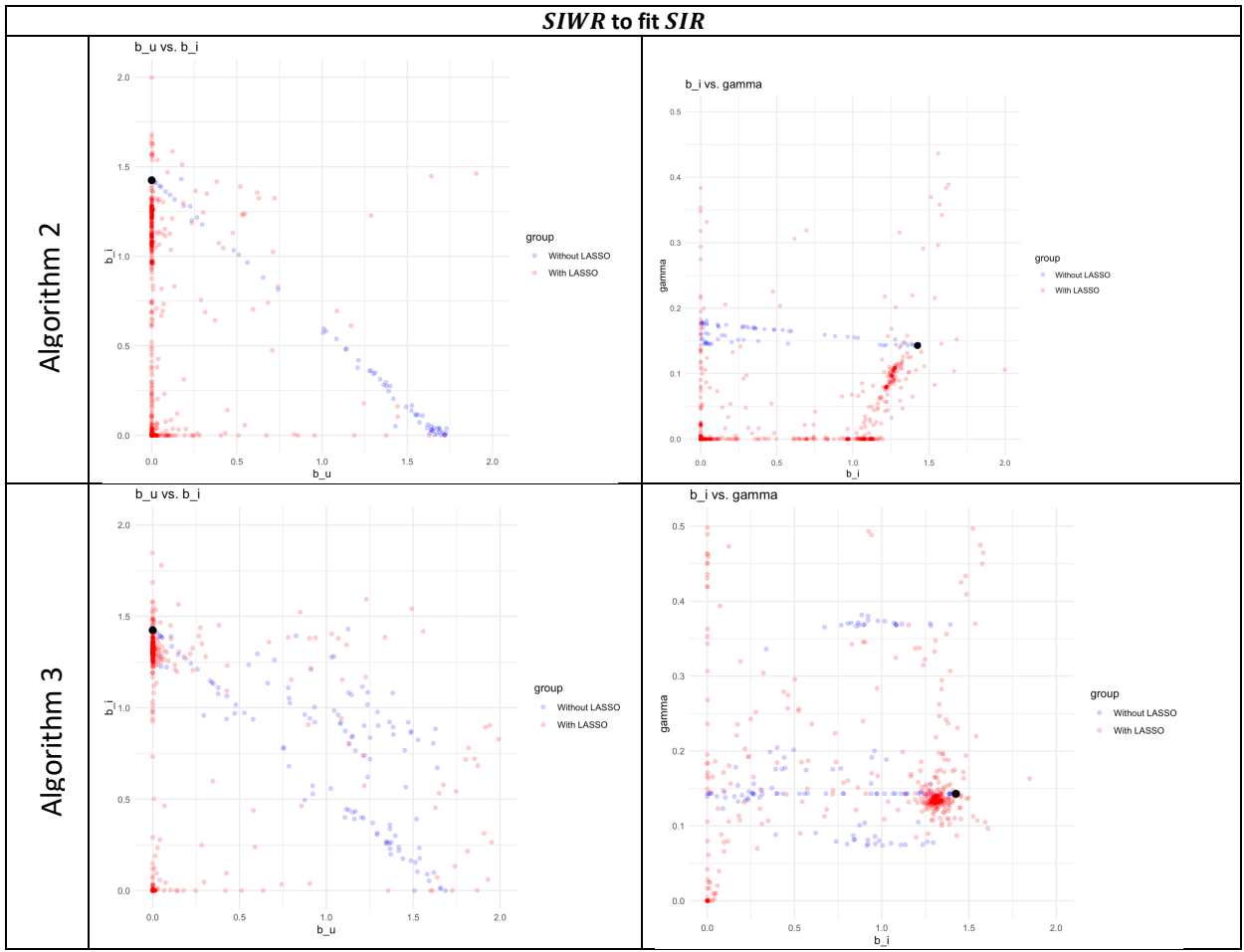

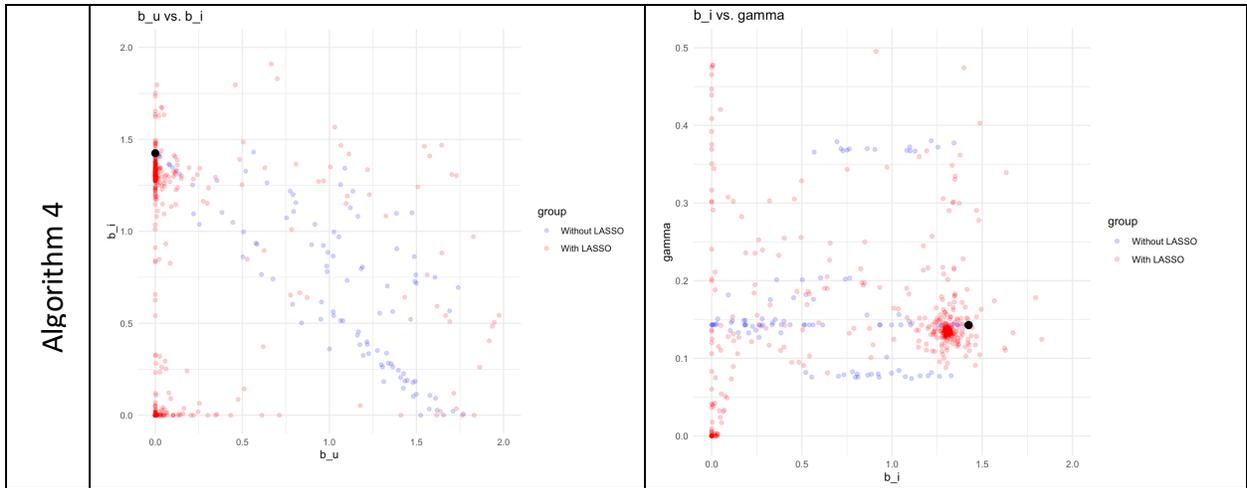

S2. Model estimates from *SIWR* to fit *SIR* scenario (Algorithm 2 - 4). Note: Here b_i, b_u, gamma are equivalent to $\beta_I$, $\beta_W$, $\gamma$ in Table 1 in the main draft.

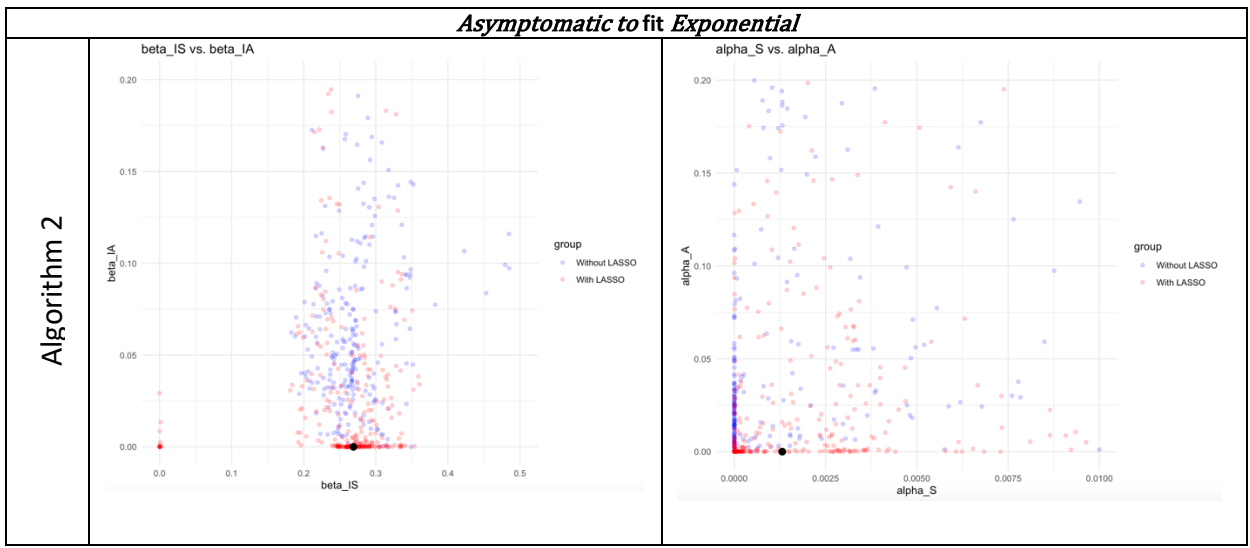

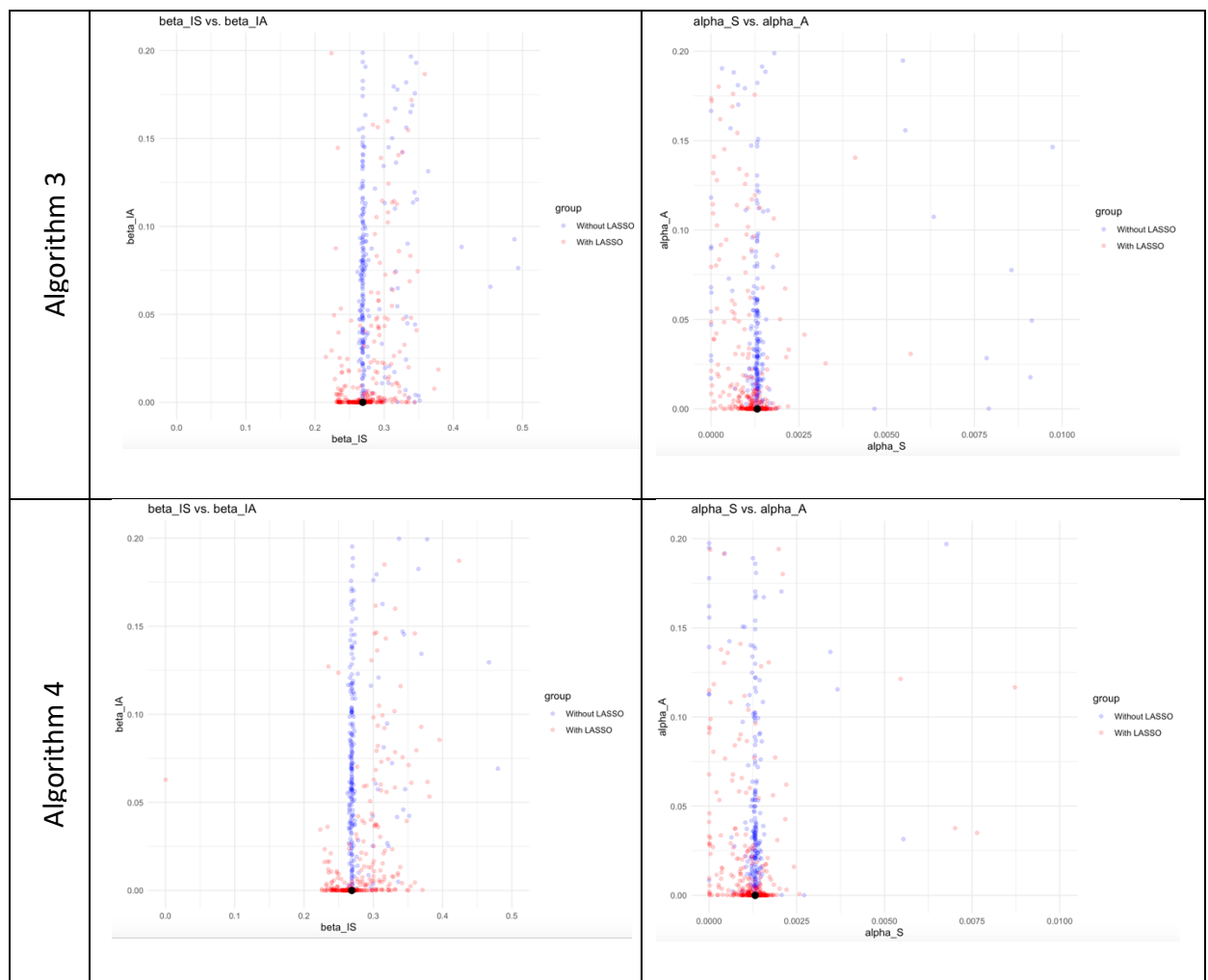

S3. Model estimates from *Asymptomatic to fit Exponential* scenario (Algorithm 2 - 4). Note: Here beta_IA, beta_IS, alpha_A, alpha_S are equivalent to $\beta_{IA}$, $\beta_{IS}$, $\alpha_A$, $\alpha_S$ in Table 1 in the main draft.

From Tables above (S1-S3), we can see that, overall, algorithms 3 and 4 perform better than algorithm 2 because the clustering dots are closer to the ground truth parameters. This can be explained by the fact that the most important data, which contains more useful information for model fitting, has been split into two parts, thus losing some of its power. Additionally, without LASSO, all the estimations (blue dots) align linearly, resulting in non-unique estimations. However, adding a LASSO penalty makes our estimations more identifiable. It is also worth noting that algorithms 3 and 4 took longer to run compared to algorithms 1 and 2 due to the larger training and testing sample sizes. In addition, Table S1 showed us all four algorithms performed better than conventional CV.

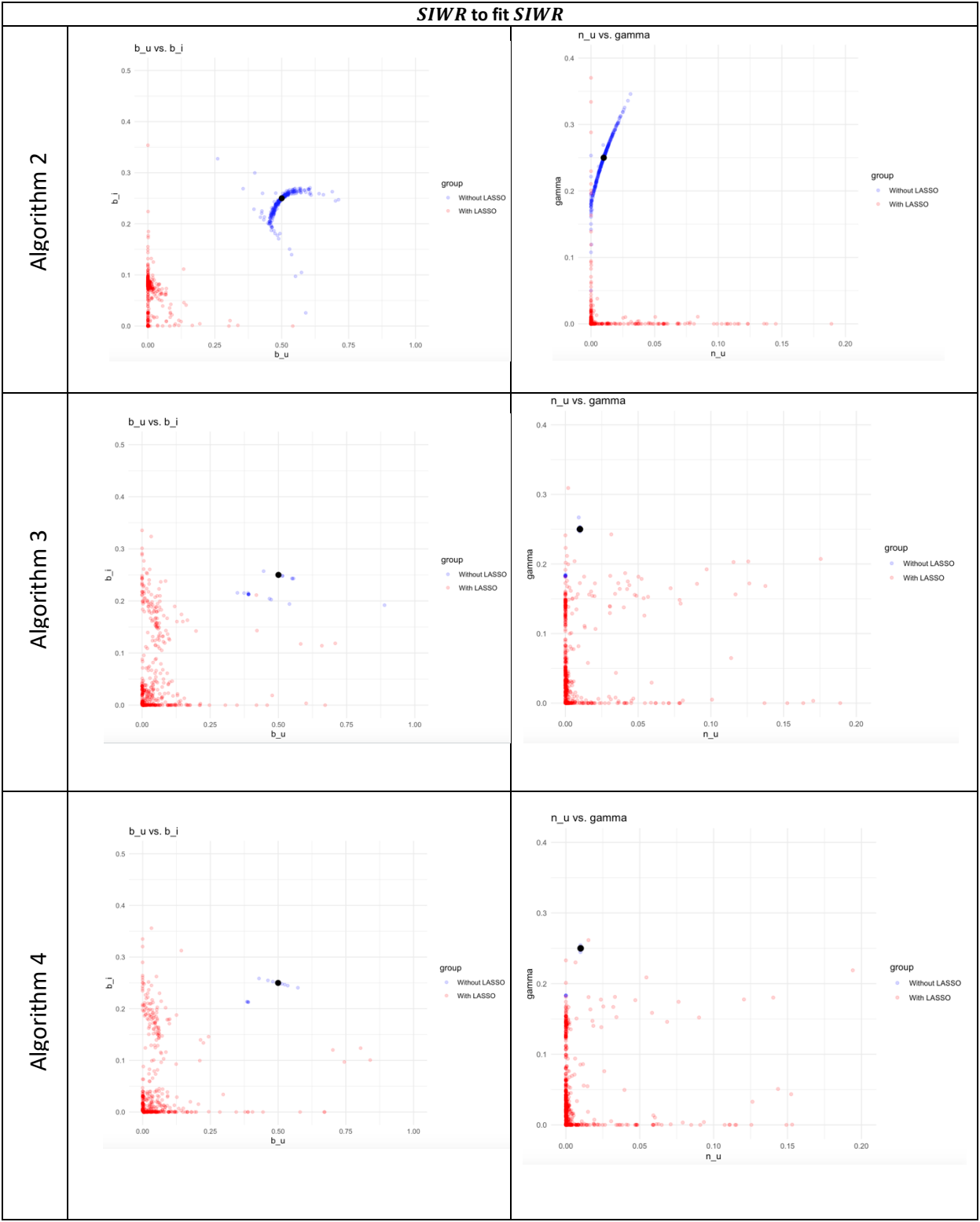

S4. Model estimates from $SIWR$ to fit $SIWR$ scenario (Algorithm 2 - 4). Note: Here b_i, b_u, n_u, gamma are equivalent to $\beta_I$ , $\beta_W$, $\xi$ , $\gamma$  in Table 1 in the main draft.

From table S4, we observe that, in general, all algorithms indicate that the best fit does not require a LASSO penalty. Specifically, Algorithms 3 and 4 provided highly accurate estimations without the LASSO penalty, as all their estimates are very close to the ground truth. This aligns with our expectations, as there is no need to simplify our model. In other words, our framework can maintain the complexity of the model effectively.

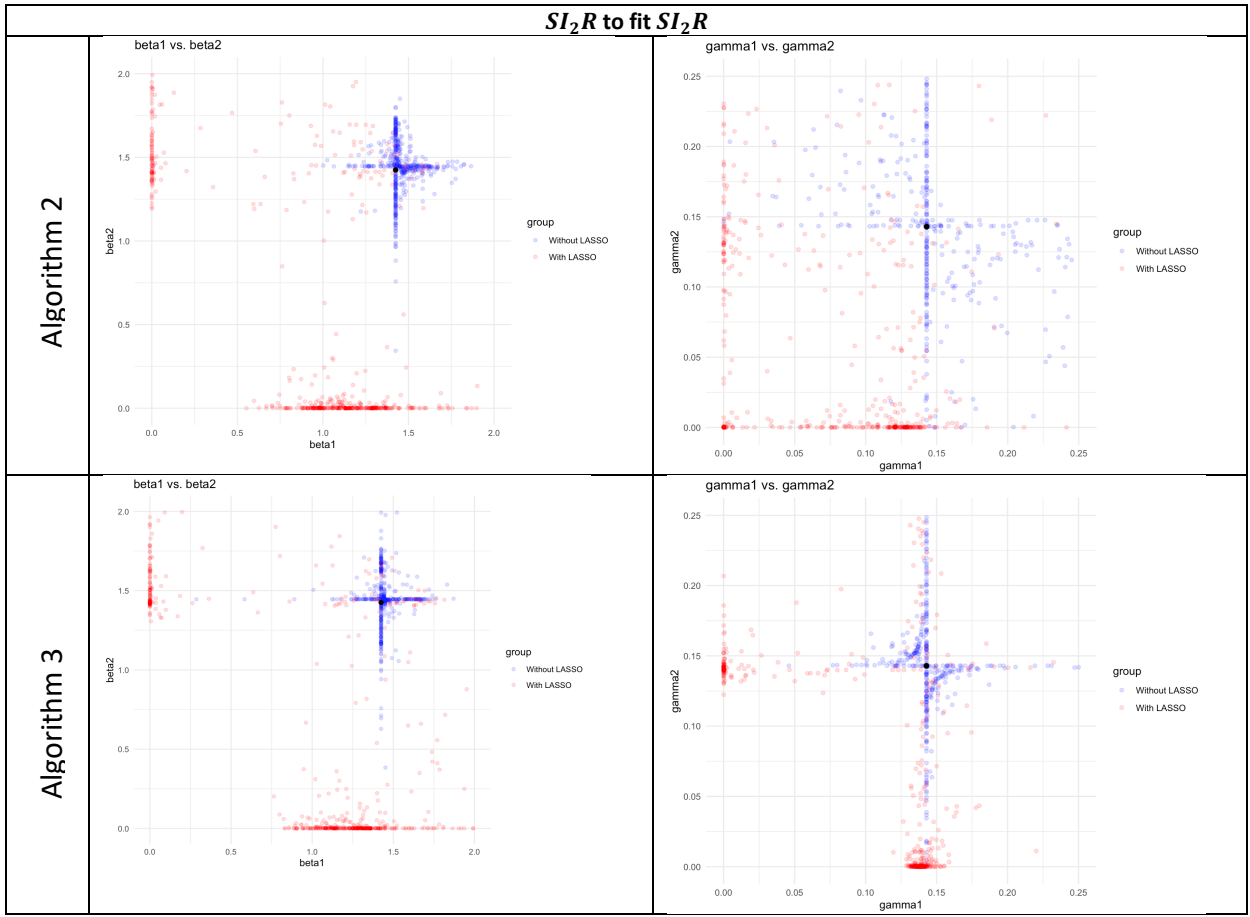

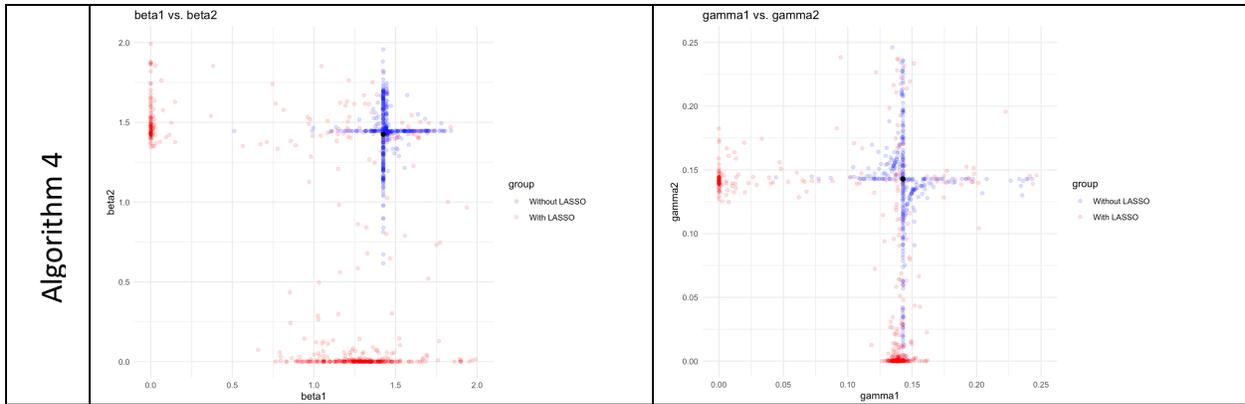

S5. Model estimates from $SI_2R$ to fit $SI_2R$ scenario (Algorithm 2 - 4). Note: Here beta1, beta2, gamma1, gamm2 are equivalent to $\beta_1$, $\beta_2$, $\gamma_1$, $\gamma_2$ in Table 1 in the main draft.

From Table S5, we observe that the ground truth parameters lie on the blue line (representing the model without LASSO). However, the Mean Squared Error (MSE) from Table S9 indicates that the fit obtained with the LASSO penalty is comparable to the fit without LASSO. Consequently, our framework may incorrectly favor the simpler model. This highlights a limitation of our approach: when the true model is highly unidentifiable, we cannot achieve accurate parameter estimation.

**Supplementary Materials Section 3: Mean-squared error results for all cross-validation algorithms from simulated data.**

The tables below (S6-S11) present the mean square error (MSE) on testing datasets, calculated using parameters estimated from training datasets. This aids in selecting our optimal lambda. For each lambda, we repeatedly generate training and testing datasets using different algorithms, resulting in a distribution of MSE values for the testing datasets. The optimal lambda is chosen based on the distribution of these MSE values.

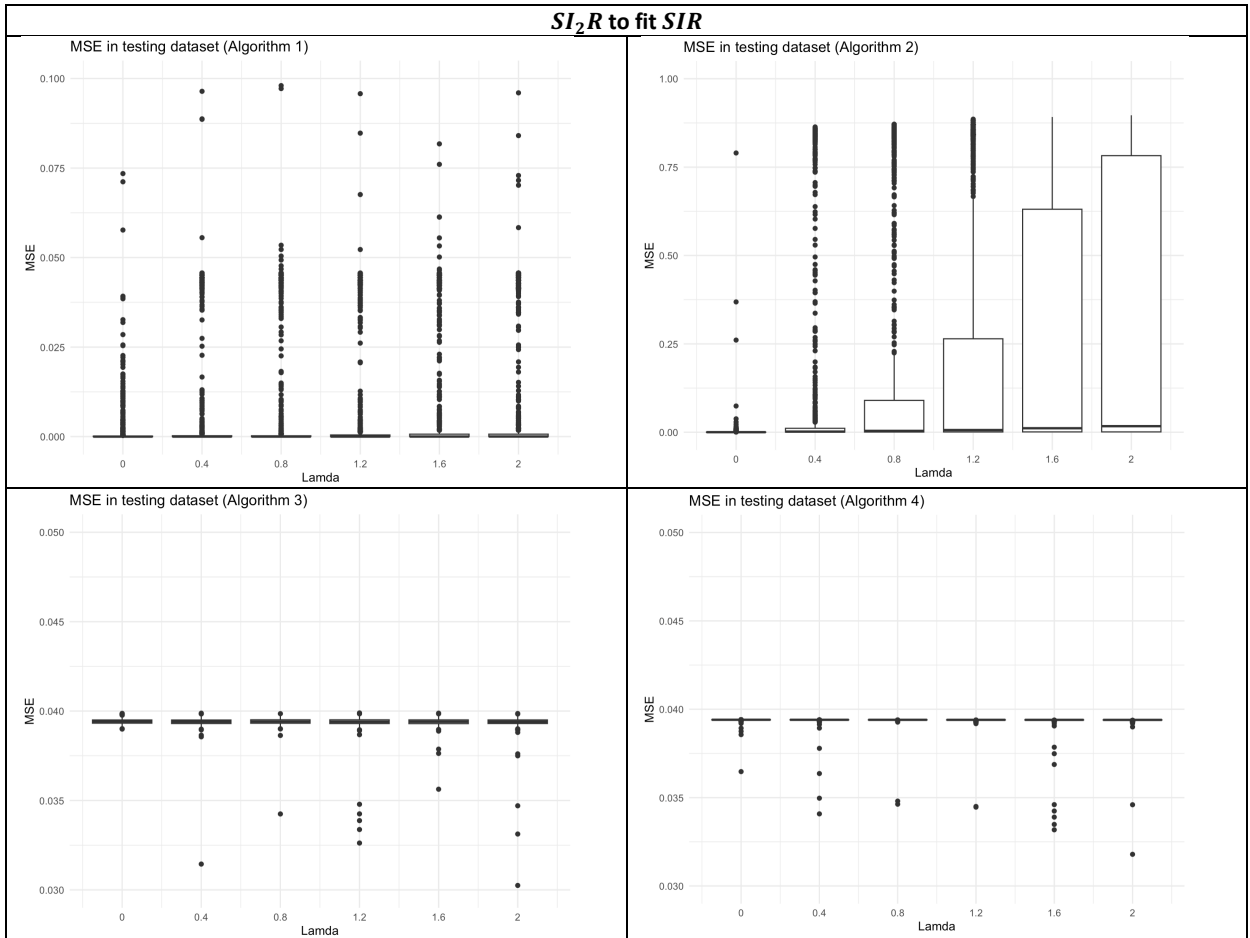

S6. MSE from $SI_2R$ to fit $SIR$ scenario (Algorithm 1 - 4)

From the table above, we observe that Algorithms 1, 3 and 4 successfully show no difference in model fit with or without lambda. This finding is consistent with the fact that the SI2R model is non-identifiable, allowing for multiple unique solutions. However, incorporating lambda provides a more simplified estimation (closer to the ground truth), demonstrating that LASSO has the capability to identify simplified results from among all possible options.

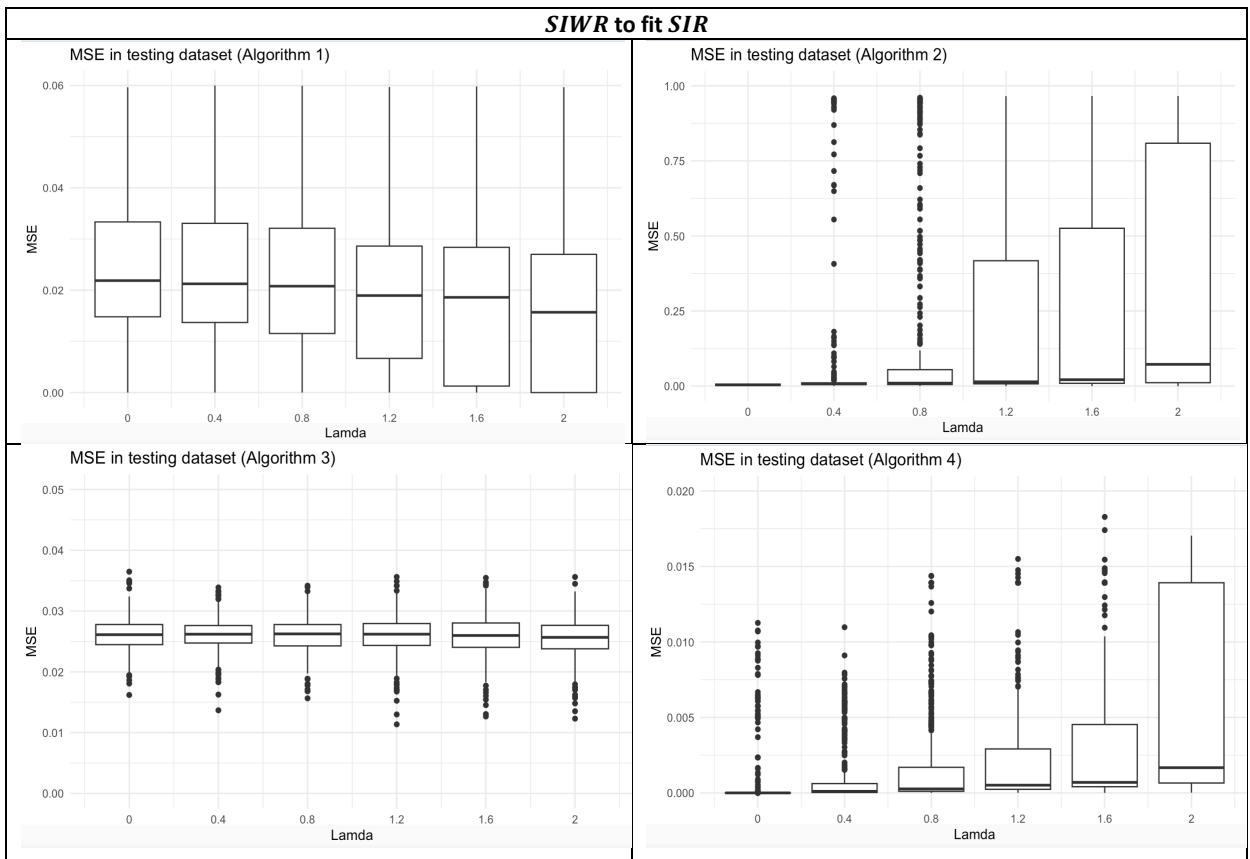

S7. MSE from $SIWR$ to fit $SIR$ scenario (Algorithm 1 - 4)

From the table above, we can see that Algorithm 1 successfully indicates that a lambda of two yields the minimum MSE. However, Algorithms 2 and 4 fail to demonstrate the effectiveness of adding lambda. Notably, the model estimates do not vary significantly among the positive lambda values we selected, indicating that our framework provides very stable estimates. In other words, as long as we add a reasonable positive lambda to the objective function, it consistently produces accurate estimates.

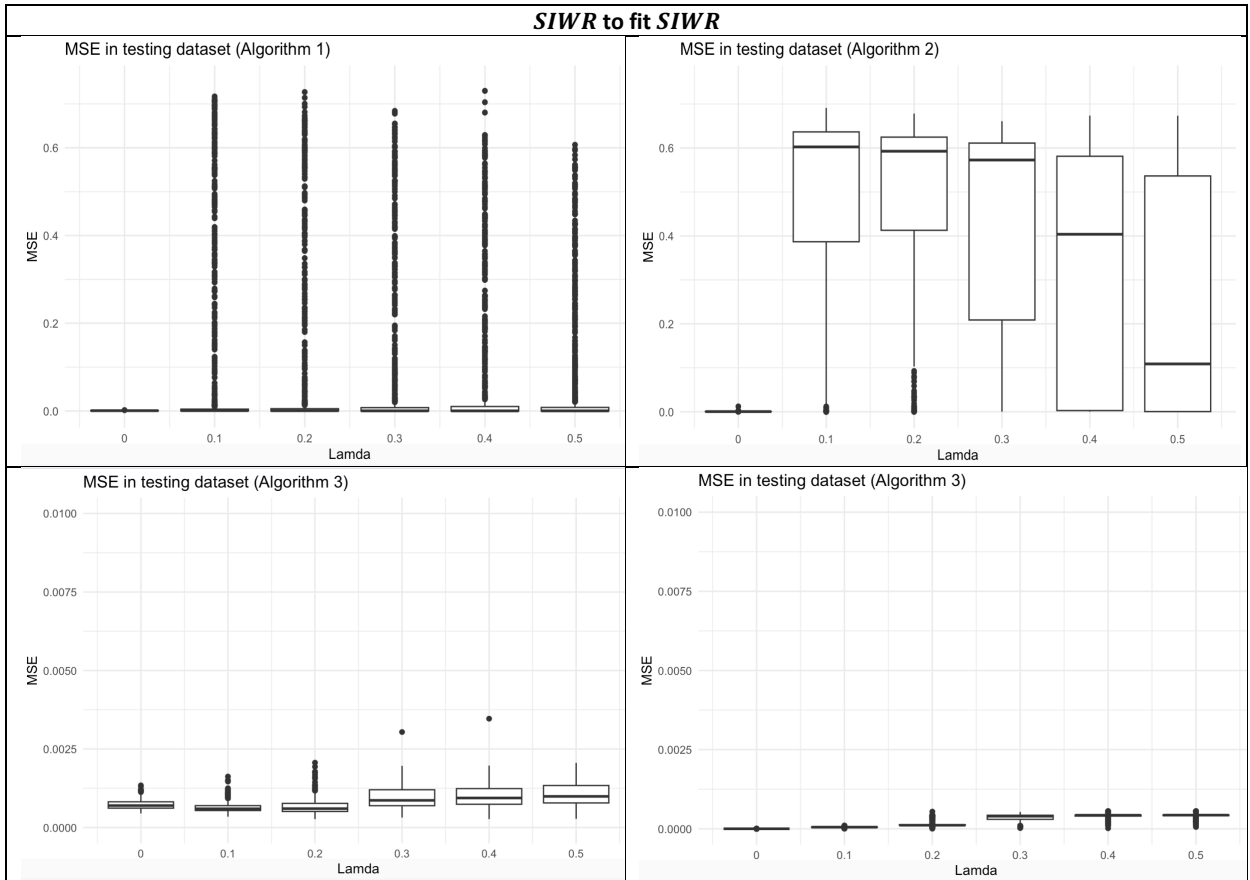

S8. MSE from $SIWR$ to fit $SIWR$ scenario (Algorithm 1 - 4)

From the table above, we can see that all algorithms successfully demonstrate that there is no need to include lambda in the objective function. In other words, we do not need to shrink any parameters in the model. This aligns with our expectations, as simplifying the model is unnecessary when fitting the model to itself. Furthermore, it shows that our framework has the capability to preserve the complexity of the model, in addition to simplifying it when required.

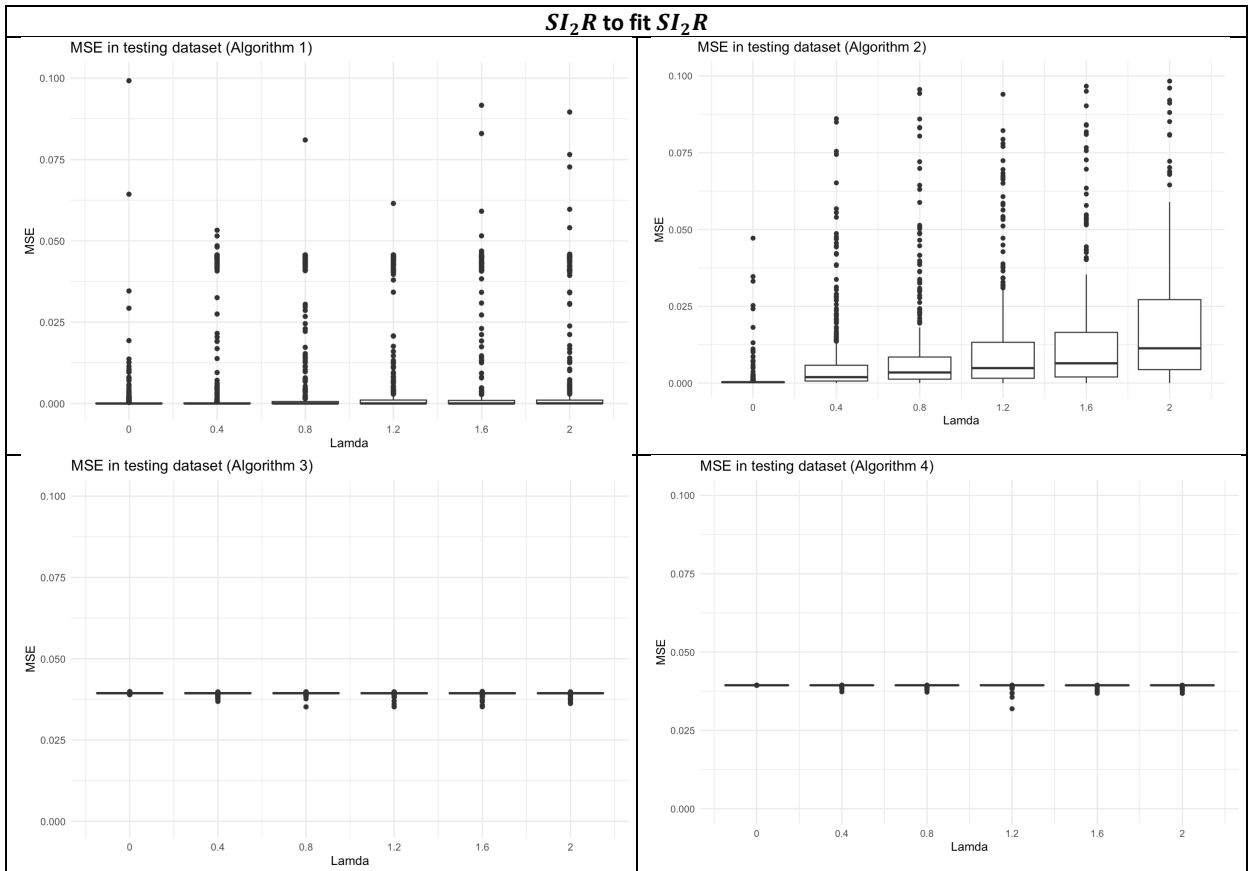

S9. MSE from $SI_2R$ to fit $SI_2R$ scenario (Algorithm 1 - 4)

Since the true model SI₂R is highly structurally unidentifiable, and LASSO-ODE in Algorithm1,3,4 show no difference in model fit with or without lambda. In other words, we misleadingly chose the simplified version of model (lambda > 0). This is the only failed case in our study, indicates that if the true model is highly structurally non-identifiable (such as the SI₂R model), our framework may fail to achieve the correct estimation

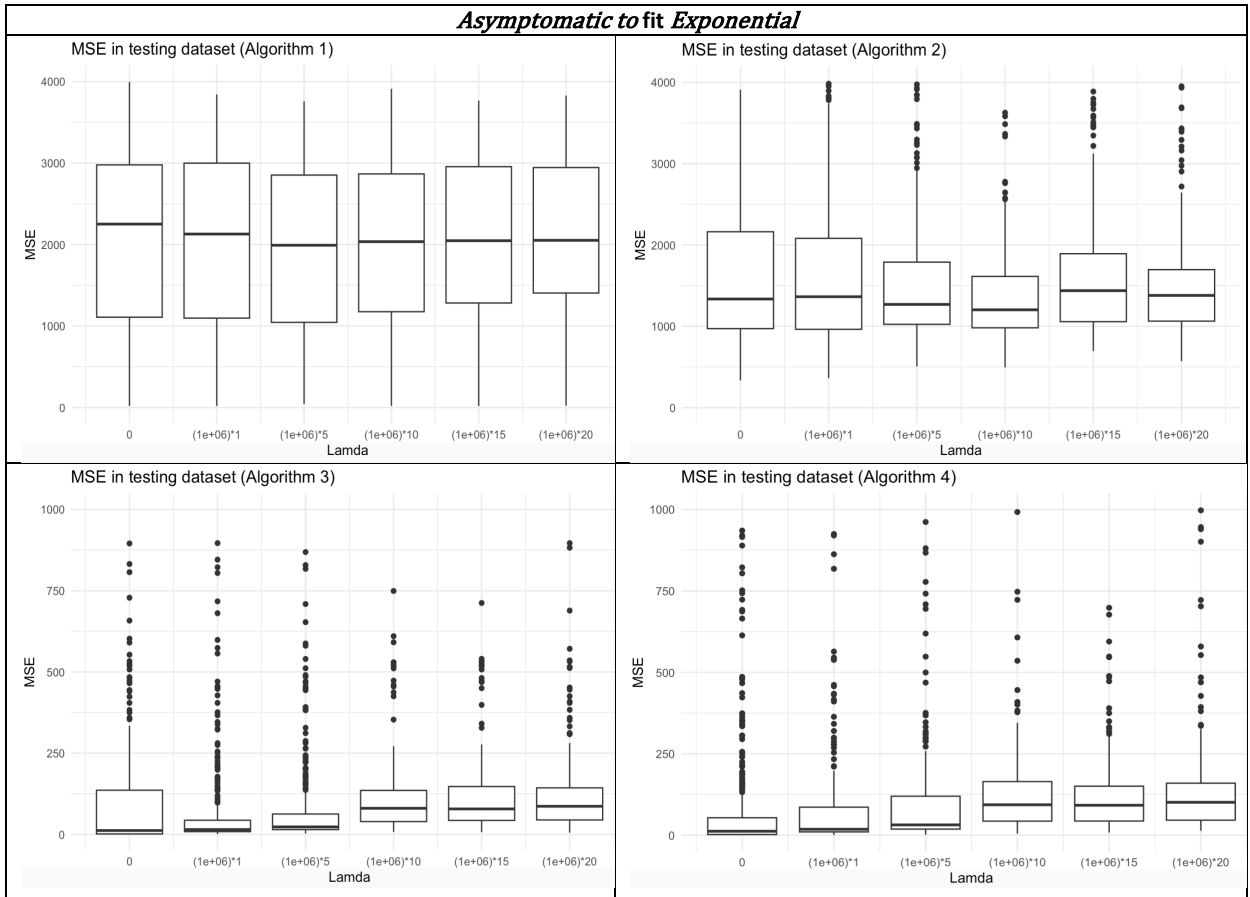

S10. MSE from Asymptomatic to fit Exponential scenario (Algorithm 1 - 4)

**Supplementary Materials Section 4: Mean-squared error results from real-world data.**

Real-world Data model fit

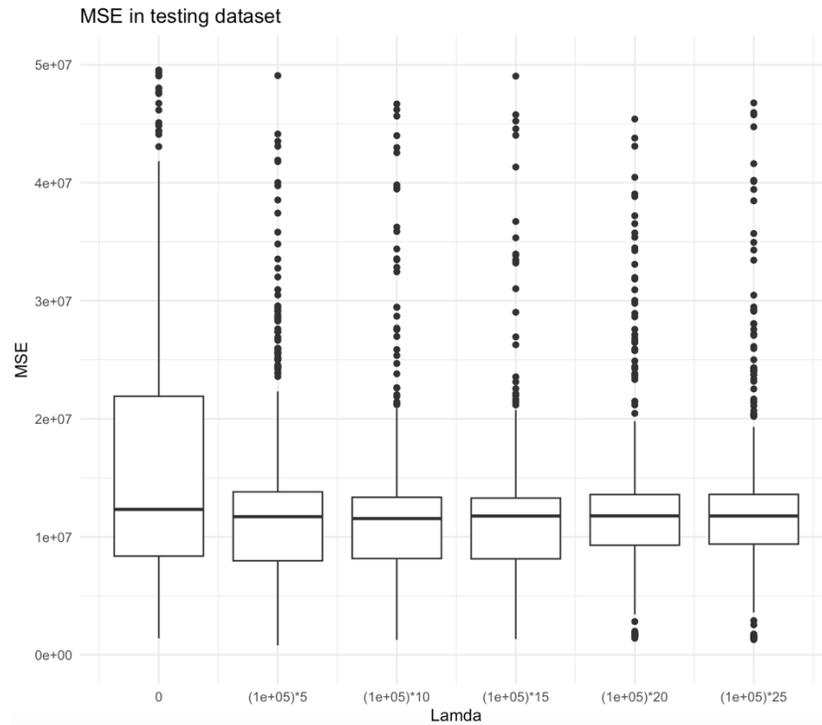

S11. MSE from real-world data scenario 1 (Algorithm 1)

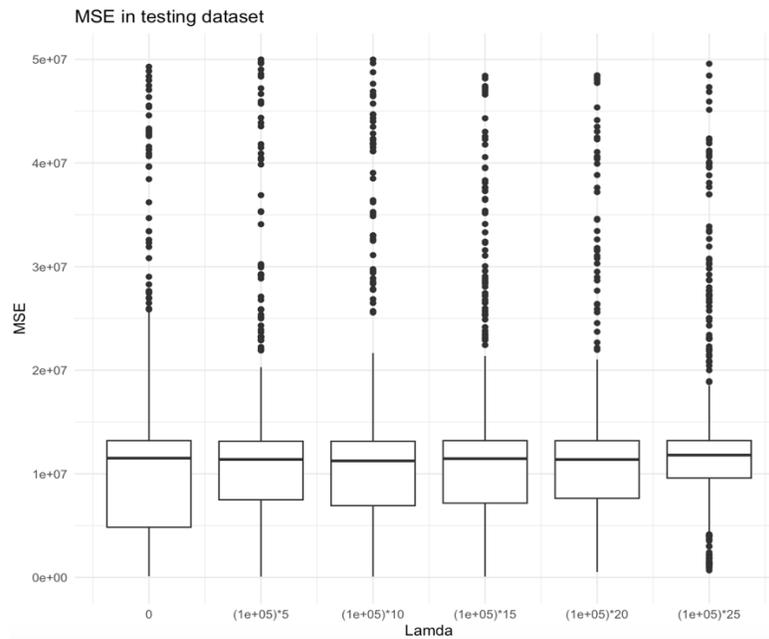

S12. MSE from real-world data scenario 2 (Algorithm 1)

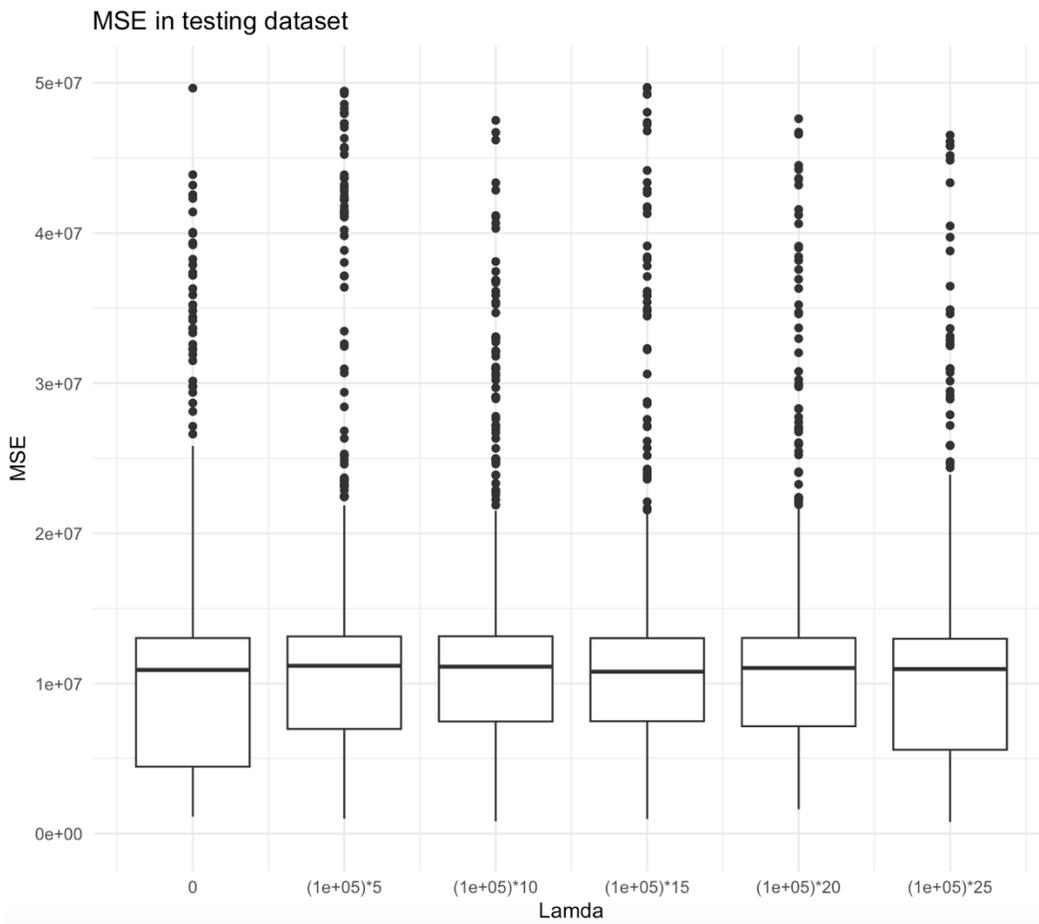

S13. MSE from real-world data scenario 3 (Algorithm 1)

S11 demonstrate that when lambda is greater than 0, it results in optimal estimation. On the other hand, S12 and S13 shows no significant difference between using lambda and not using it. However, our reasoning is to always choose the simplified version when the mean squared error (MSE) is similar.

**Supplementary Materials Section 5: Model fits from simulated data.**

The following figures (S14-S19) show the time series plots with data and the corresponding predictions from our LASSO-ODE framework:

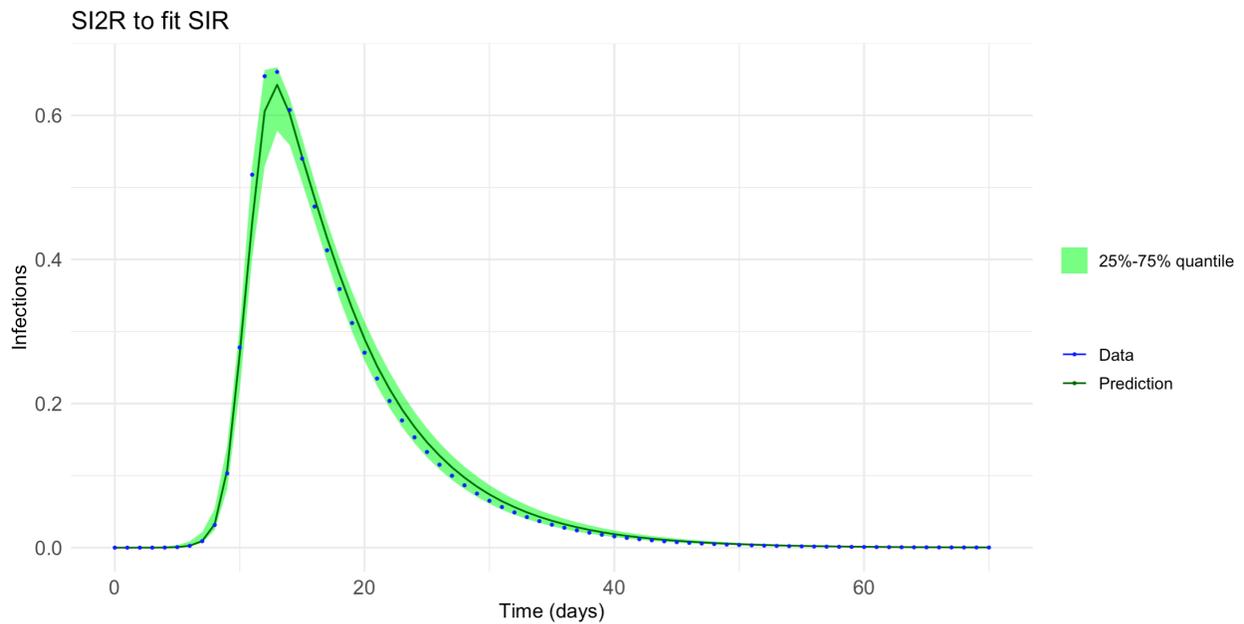

S14. Time series plot from SI2R to SIR data scenario (Algorithm 1)

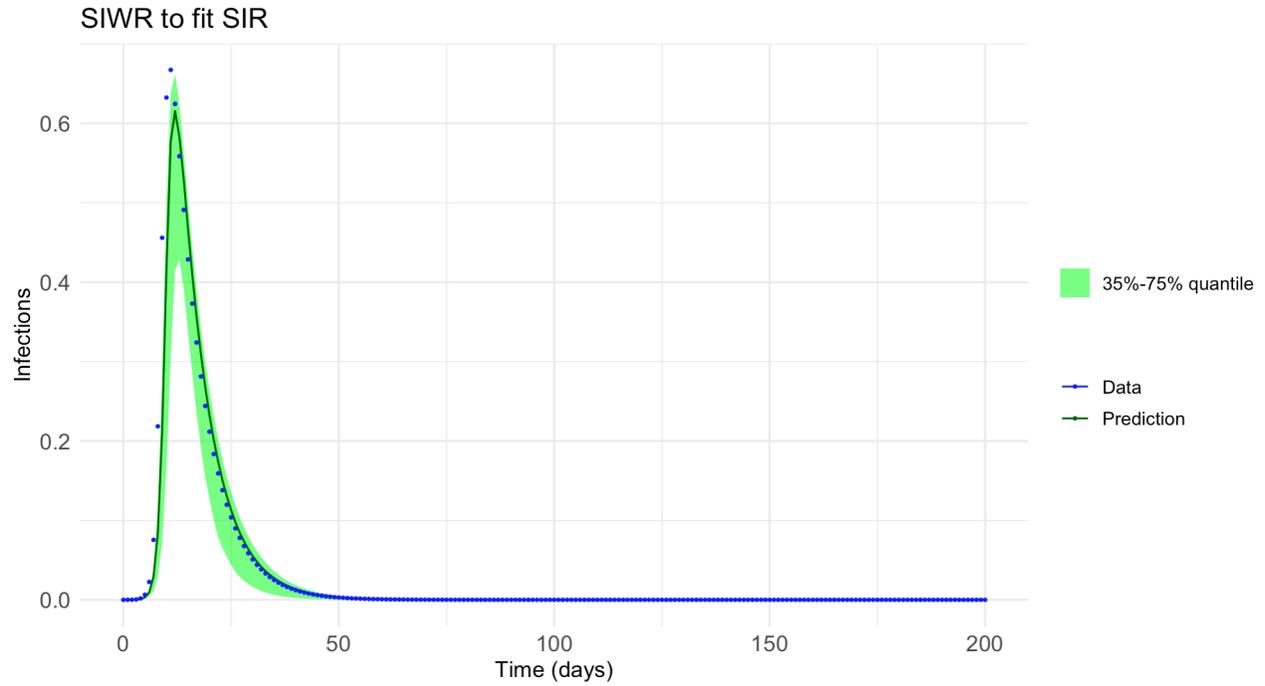

S15. Time series plot from SIWR to SIR data scenario (Algorithm 1)

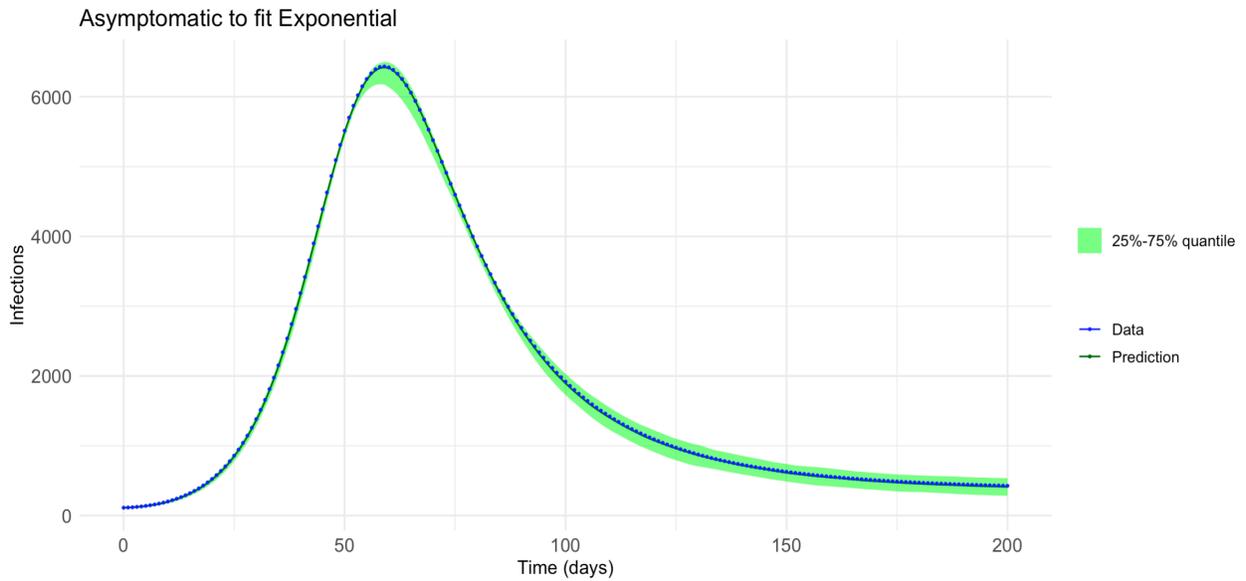

S16. Time series plot from asymptomatic to exponential data scenario (Algorithm 1)

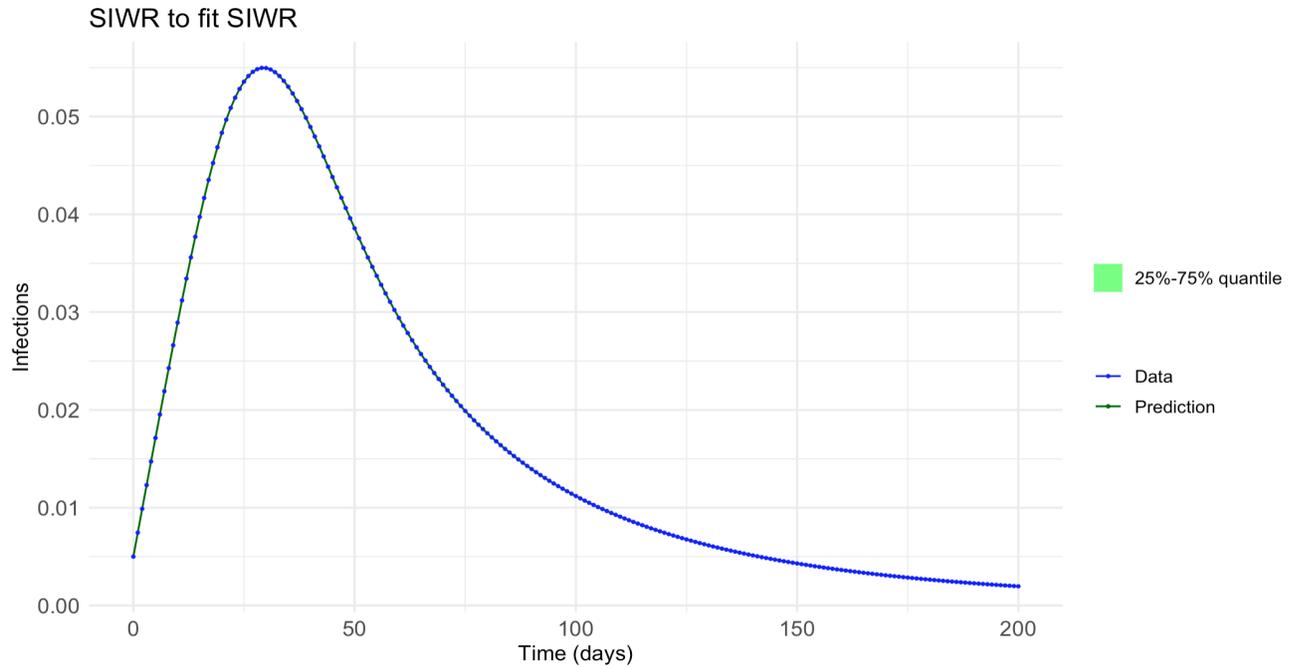

S17. Time series plot from SIWR to SIWR data scenario (Algorithm 1)

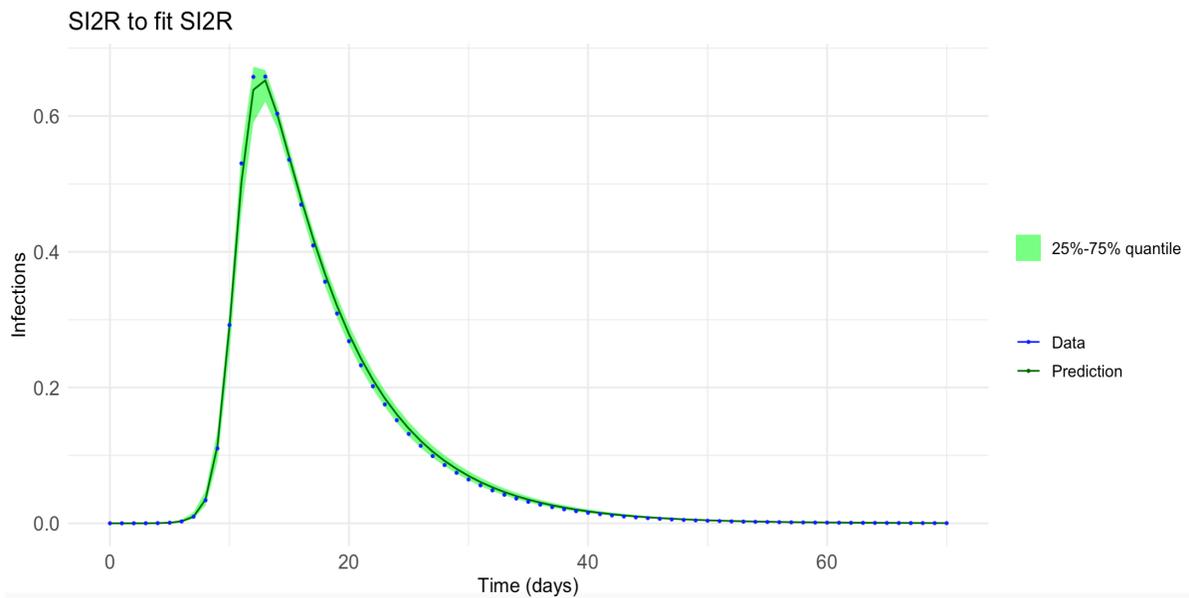

S18. Time series plot from SI2R to SI2R data scenario (Algorithm 1)

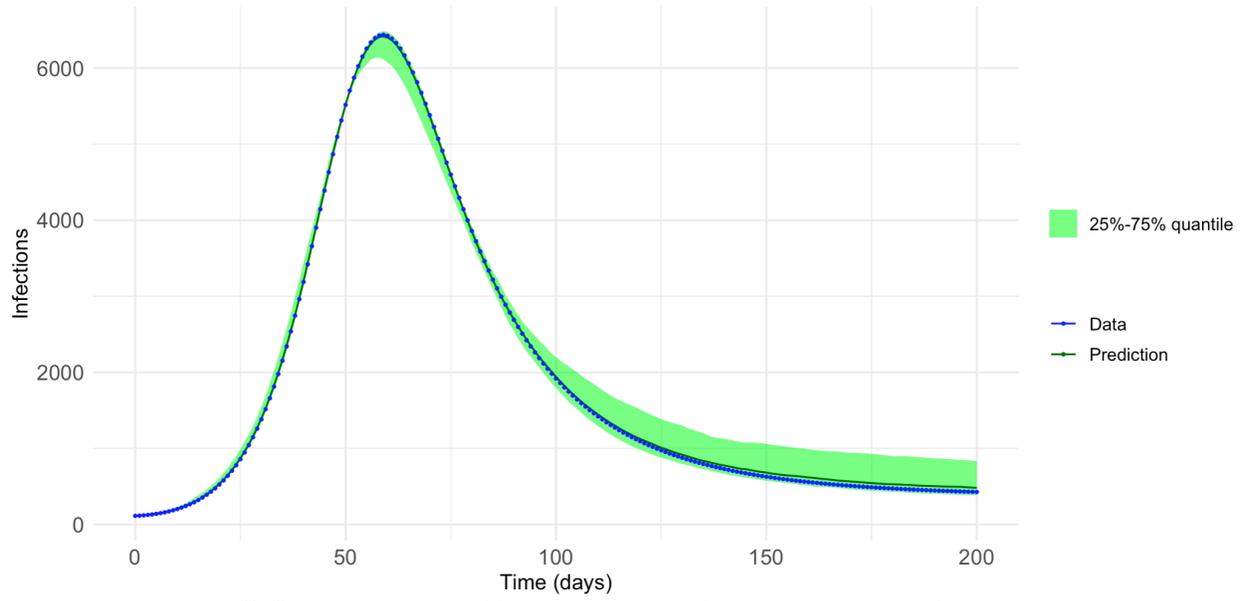

S19. Time series plot from model selection scenario (Algorithm 1)

**Supplementary Materials Section 6: Parameter estimates from simulated data.**

The following tables (S20-S26) show the distribution (median [1st quantile ,3rd quantile]) of all estimated parameters of each scenario from our LASSO-ODE framework:

| SI2R to SIR | | | |
|---|---|---|---|
| Parameters | True value | With LASSO penalty (Optimal) | Without LASSO penalty |
| $\beta_1$ | 1.4247 | 1.0500 [0.7391, 1.2594] | 1.425 [1.425, 1.425] |
| $\beta_2$ | 0 | 0.0002353 [0.0000027, 0.0129461] | 0.35084 [0.15351, 0.71253] |
| $\gamma_1$ | 0.14286 | 0.13115 [0.08535, 0.13752] | 0.142860 [0.142856, 0.142866] |
| $\gamma_2$ | 0 | 0.0003146 [0.0000051, 0.0178621] | 0.193074 [0.077073, 0.510423] |
| $I_1$ | - | 2.321e-05 [3.650e-06, 2.655e-04] | 1.154e-06 [1.153e-06, 1.155e-06] |

S20. All parameters in the SI2R model

| SIWR to SIR | | | |
|---|---|---|---|
| Parameters | True value | With LASSO penalty(Optimal) | Without LASSO penalty |
| $\beta_w$ | 0 | 0.000561 [0.000001, 0.076718] | 0.0001121 [0.0000345, 0.0046634] |
| $\beta_I$ | 1.4247 | 1.2821 [0.1096, 1.3181] | 1.431 [1.427, 1.431] |
| $\xi$ | 0 | 0.000644 [0.000000, 0.072511] | 7.905 [7.435, 8.734] |
| $\gamma$ | 0.14286 | 0.1353 [0.1242, 0.1896 ] | 0.14290 [0.14289, 0.14292] |

S21. All parameters in the SIWR model

| Asymptomatic to Exponential | | | |
|---|---|---|---|
| Parameters | True value | With LASSO penalty(Optimal) | Without LASSO penalty |
| $\beta_{IS}$ | 0.269 | 0.2689959 [0.2476271, 0.2887733] | 0.2704 [0.2669, 0.3047] |
| $\beta_{IA}$ | 0 | 0.0006978 [0.0000613, 0.0100659] | 0.094870 [0.042260, 0.218345] |
| $\beta_w$ | 1.62 | 1.7461 [1.6460, 1.8342] | 1.628282 [1.530909, 1.714496] |
| $\alpha_S$ | 0.001314 | 0.0013171 [0.0008551, 0.0018417] | 0.001386 [0.001030, 0.007106] |
| $\alpha_A$ | 0 | 0.0020420 [0.0000931, 0.0165220] | 0.05388 [0.01853, 0.20007] |
| $\xi$ | 0.00614 | 0.0061493 [0.0058506, 0.0073268] | 0.006127 [0.005751, 0.006324] |
| $k$ | 0.0000116 | 0.0000116 [0.0000106, 0.0000123] | 1.156e-05 [1.004e-05, 1.183e-05] |
| $q$ | 1 | 0.9868 [0.8949, 1.0001] | 1.0177 [0.9995, 1.1199] |

S22. All parameters in the Asymptomatic model

| SIWR to SIWR | | | |
|---|---|---|---|
| Parameters | True value | With LASSO penalty | Without LASSO penalty (Optimal) |
| $\beta_w$ | 0.5 | 0.0142951[0.0003013, 0.0486252] | 0.5000 [0.4999 , 0.5001] |
| $\beta_I$ | 0.25 | 0.0119072 [0.0002704, 0.0516629] | 0.2500 [0.2500, 0.2500] |
| $\xi$ | 0.01 | 0.0004873 [0.0000002, 0.0100245] | 0.010000 [0.009994, 0.010005] |
| $\gamma$ | 0.25 | 1.002e-03 [2.900e-07, 3.422e-02] | 0.2500 [0.2500, 0.2500] |

S23. All parameters in the SIWR model

| SI2R to SI2R | | | |
|---|---|---|---|
| Parameters | True value | With LASSO penalty (Optimal) | Without LASSO penalty |
| $\beta_1$ | 1.4247 | 1.16676 [0.04538, 1.35541] | 1.4264 [1.4246, 1.4246] |
| $\beta_2$ | 1.4247 | 0.0165081 [0.0000541, 1.4127421] | 1.4457 [1.4172, 1.5110] |
| $\gamma_1$ | 0.14286 | 0.136081 [0.008603, 0.142135] | 0.1428614[0.1427132, 0.1682950] |
| $\gamma_2$ | 0.14286 | 0.0869568 [0.0000777, 0.1444480] | 0.1429828[0.1323230, 0.1797827] |
| $I_1$ | - | 2.082e-06 [2.760e-07, 1.442e-05] | 7.976e-07 [5.006e-08, 1.248e-06] |

S24. All parameters in the SI2R model

| Model selection | | | |
|---|---|---|---|
| Parameters | True value | With LASSO penalty(Optimal) | Without LASSO penalty |
| $\beta_{IS}$ | 0 | 0.0156974[0.0013861, 0.0489889] | 0.0589026[0.0409906, 0.1072957] |
| $\beta_{IA}$ | 0 | 0.0106091[0.0017662, 0.0241269] | 0.030079[0.019135, 0.081683] |
| $\beta_{aw}$ | 0 | 0.1575646[0.0024348, 0.2856802] | 0.302587[0.266874, 0.348742] |
| $\alpha_S$ | 0 | 0.0057423[0.0008371, 0.0317821] | 0.022777[0.005141, 0.082632] |
| $\alpha_A$ | 0 | 0.0059464[0.0010085, 0.0231346] | 0.027317[0.007127, 0.091184] |
| $\xi_a$ | 0 | 0.0550855[0.0194780, 0.1232173] | 0.022472[0.006259, 0.092018] |
| $k_a$ | 0 | 0.05609[0.01310, 0.15202] | 0.027283[0.007789, 0.085008] |
| $q$ | 1 | 1.0324[1.0000, 1.1024] | 1.0098[0.9999, 1.0486] |
| $\beta_I$ | 0.269 | 0.2631884[0.2284086, 0.2842513] | 0.2698[0.2587, 0.2960] |
| $\beta_w$ | 1.62 | 1.6869[1.5101, 1.8435] | 1.6308728[1.4501837, 1.8140011] |
| $\xi$ | 0.00614 | 0.0061536[0.0060848, 0.0071993] | 0.0061400[0.0060862, 0.0061829] |
| $\alpha$ | 0.001314 | 1.455e-03[1.049e-03, 2.258e-03] | 0.001476[0.001132, 0.003054] |
| $k$ | 0.0000116 | 0.0000112[0.0000090, 0.0000123] | 1.121e-05[9.321e-06, 1.171e-05] |
| w | - | 0.0879668[0.0240428, 0.1879871] | 0.028474[0.008904, 0.111932] |

S25. All parameters in the model selection